\documentclass[11pt,doublespace,onecolumn]{IEEEtran}

\usepackage[dvips]{epsfig}
\usepackage[dvips]{graphicx}
\usepackage{boxedminipage}
\usepackage{color}
\usepackage{amsfonts}
\usepackage{amssymb}
\usepackage{amsmath}
\usepackage{nopageno}
\usepackage{graphics,amsmath,amsfonts,amssymb,epsfig,latexsym,amsfonts,graphicx,amssymb}
\usepackage{setspace}
\usepackage{cite}

   

\usepackage{eqparbox}

\newcommand{\bfbeta}{{\mbox{\boldmath $\beta$}}}

\newcommand{\bx}{\mathbf{x}}

\newcommand{\ba}{\mathbf{a}}
\newcommand{\bp}{\mathbf{p}}

\makeatletter
\let\sv@endpart\@endpart
\def\@endpart{\thispagestyle{empty}\sv@endpart}
\makeatother

\begin{document}
\title{Mechanism Design for Base Station Association and Resource Allocation in Downlink
OFDMA Network} \vspace{-0.1cm}
\author{Mingyi Hong and Alfredo Garcia
\thanks{\footnotesize M. Hong was with the Department of Systems and Information Engineering,
University of Virginia, and is now with the OSPAC group, the Digital
Technology Center (DTC), Department of Electrical and Computer
Engineering, University of Minnesota. A. Garcia is with the
Department of Systems and Information Engineering, University of
Virginia. } }

\maketitle \vspace{-0.3cm}
\begin{abstract}
\vspace{-0.3cm}

We consider a resource management problem in a multi-cell downlink
OFDMA network whereby the goal is to find the optimal combination of
{\em (i)} assignment of users to base stations and {\em (ii)}
resource allocation strategies at each base station. Efficient
resource management protocols must rely on users {\em truthfully}
reporting privately held information such as downlink channel
states. However, individual users can manipulate the resulting
resource allocation (by misreporting their private information) if
by doing so can improve their payoff. Therefore, it is of interest
to design efficient resource management protocols that are {\em
strategy-proof}, i.e. it is in the users' best interests to
truthfully report their private information. Unfortunately, we show
that the implementation of any protocol that is efficient {\em and}
strategy-proof is NP-hard. Thus, we propose a computationally
tractable strategy-proof mechanism that is {\em approximately}
efficient, i.e. the solution obtained yields at least $\frac{1}{2}$
of the optimal throughput. Simulations are provided to illustrate
the effectiveness of the proposed mechanism.

\end{abstract}
\begin{IEEEkeywords}
Heterogenous Network, Mechanism Design, Resource Allocation, Base
Station Association, Approximation Bounds, Computational Complexity,
Nash Equilibrium, Price of Anarchy
\end{IEEEkeywords}

\vspace{-0.4cm}
\section{Introduction}

We consider a downlink OFDMA network with multiple base stations
(BSs) serving a group of users. The BSs operate on non-overlapping
spectrum bands in frequency division duplex (FDD) mode. The
objective is to find the best per-BS resource allocation strategy
and the user-BS assignment to achieve spectral efficiency and load
balancing across the networks. This problem is well motivated by
many practical networks such as the multi-technology heterogenous
networks (HetNet) \cite{McNair04, yeh11}, the IEEE 802.22 Wireless
Regional Area Network (WRAN) \cite{fcc10} or a Wi-Fi network with
multiple access points \cite{kauffmann07}. For example, in the
HetNet, multiple wireless access technologies such as Wi-Fi, LTE or
WiMAX are available for the same region. These networks operate on
different spectrum bands and all utilize OFDMA for downlink
transmission. Integrating these different radio access technologies
and making them available to the user devices can significantly
increase the overall spectral efficiency as well as achieve load
balancing across different networks. The mobile users can choose
from one of the technologies/networks for association, and they can
switch between different technologies/networks to avoid congestion
(i.e., ``vertical handoff" operation, see \cite{McNair04}). The
user-network assignment and the per-network resource allocation need
to be performed jointly to achieve optimal network-wide resource
allocation.

There are three major challenges for optimal resource allocation in
such networks.

1) When operating in FDD mode, the network requires the users to
measure and report the downlink channel states for efficient
resource allocation. An untruthful user may incorrectly report this
information for its own benefit. The possibilities of various forms
of untruthfulness in user behaviors in wireless networks have been
recently noted (see e.g.,
\cite{Kavitha12,Bianchi07,Nuggehalli08,Kong07}). Reference
\cite{Bianchi07} has discovered that certain commercial 802.11
devices are specially manipulated to trick the access points for
better rates. Such manipulation can take the form of a non-uniform
backoff procedure \cite{Bianchi07} or falsely reported traffic
priorities \cite{Nuggehalli08}. As suggested in \cite{Kavitha12}, in
FDD cellular networks it is also possible to manipulate the devices'
channel feedback procedure, as the compliance testing is usually
limited to a few standardized scenarios. It is projected that the
users of future networks may have stronger ability (and incentives)
for manipulating their devices. See \cite{Kavitha12} for detailed
literature review in this direction. The presence of the untruthful
users can significantly reduce the overall system performance and
limit network access to truthful users.

2) Even assuming the users truthfully report their channel states,
finding the global optimal resource allocation is still
computationally intractable (this result will be shown in Section
\ref{secSystemModel}).

3) There is no central entity to compute and enforce a desired
user-network assignment and network-wide resource allocation
\cite{kauffmann07}.

Consequently, a good resource allocation scheme must possess the
following features: {\em i)} it should provide efficient utilization
of the spectrum; {\em ii)} it must be {\em strategy-proof}, i.e., it
is in the users' best interests to truthfully reveal their private
information; {\em iii)} it is distributedly implementable, in the
sense that both the BSs and the users can take part in the scheme
with only local information and local computation.

\vspace{-0.5cm}
\subsection{Literature Review}
The problem of finding an optimal resource allocation for OFDMA
network when the user-BS assignment is {\em fixed} and there is {\em
complete} information on channel states has been widely studied,
e.g., \cite{song05a,song05b,tse_m}. Various algorithms are developed
to solve the BS's utility maximization problems by allocating the
transmission power and the channels in optimal ways. The {joint}
problem of BS assignment and resource allocation in OFDMA network
has been analyzed under {\it complete information} and the ability
to enforce decisions from a centralized standpoint, for
example, \cite{han04,Li06}. 
However, in many practical networks there are no entities capable of
performing the centralized decision making. Another strand of the
literature deals precisely with this case by using non-cooperative
game theory \cite{saraydar01,hong11_infocom,gao11, hong12_icassp}.
Users selfishly compute their power control and cell site selection
strategies to maximize their own utilities. With proper design of
the utility functions, equilibrium solutions can be obtained in
distributed fashion. However, complete information on the channel
states and/or the utility functions is assumed. The overall
efficiencies of the identified equilibrium solutions are not
characterized.

There are many recent works that design mechanisms for resource
allocation problems in  networks with strategic users and/or
incomplete information \cite{Kavitha12,Yang07,jain10}. It is
commonly assumed that resources are divisible and that there is a
closed-form expression that describes the interdependency of users'
decisions. In contrast, in our problem the resource allocation
decisions are {\it mixed} in nature, i.e., decisions such as user-BS
assignment and channel assignment are discrete (not divisible),
while the BSs' power allocation decision is continuous (divisible).
In addition, the interdependency in users' decisions is only {\it
implicitly} characterized as the solution to the optimal resource
management problem at each BS. As a result, the problem considered
in this paper does not adequately fit into any of the frameworks
considered in the above cited papers. We mention that the recent
work \cite{Kavitha12} considered an incomplete information setting
similar to ours, in which the FDD network lacks the true channel
states due to the false report by the users. The objective though is
to design optimal user scheduling algorithms, which is different
from the objective of the present paper.

Lower bounds of the efficiency of the Nash Equilibrium (NE), or the
price of anarchy, have been analyzed for network resource allocation
games. Reference \cite{Roughgarden00howbad} considered a routing
game in which the inefficiency is due to the selfishness of the
users. Reference \cite{Johari:2004} analyzed a network utility
maximization problem in which the strategic behavior of the users
leads to efficiency loss. For both of the above cases the optimal
system level problems can be solved globally, whereas in our case
the overall problem is already difficult to solve. In
\cite{vetta02}, Vetta discussed the lower bounds of the NEs for a
family of non-cooperative games {\it assuming} a special structure
of the users' utility functions. Applications of this latter result
in communication and sensor networks include \cite{ai08} and
\cite{Goemans06}. However, these works use highly stylized  utility
functions so that the result in \cite{vetta02} can be directly used.


\vspace{-0.3cm}
\subsection{Contributions}
The main contributions of our paper are summarized as
follows.
\begin{itemize}
\item We show that solving the joint BS assignment and resource allocation
problem, even with truthful users, is NP-hard. In most existing
complexity results for resource management in wireless
communications (e.g., \cite{luo08a,Razaviyayn11Asilomar}), the
hardness of the problem is mainly due to the possibility of strong
interference among the users. In contrast, in our problem the
hardness lies in its mixed (discrete and continuous) formulation.

\item  The complexity of solving the joint problem optimally
prevents the implementation of any mechanism that is both efficient and strategy-proof. To ensure
tractability, we design a novel mechanism that implements an {\em
approximately} optimal strategy. 
In the proposed mechanism each user dynamically selects a BS to
maximize its own utility, and the BSs implement the celebrated
Vickrey-Clark-Groves mechanism (VCG) \cite{clark71,
groves73,Vickrey61}  to allocate resource under the current user-BS
assignment profile.

\item We show that our mechanism achieves at least $\frac{1}{2}$ of the optimal
throughput. This result relies on a key observation that the per-BS
optimal throughput admits certain submodular property. Importantly,
the obtained efficiency bounds hold true {\it with or without} the
presence of the untruthful users. To the best of our knowledge,
there has been no reported work that characterizes the efficiency of
the equilibrium solutions for the considered problem.
\end{itemize}


The rest of the paper is organized as follows. Section
\ref{secSystemModel} formulates the problem and provides its
complexity status. Section \ref{secGame} and \ref{secAlgorithm}
describe the mechanism for the resource management problem as well
as its distributed implementation. Section \ref{secDiscussion} gives
some extensions of the algorithm. Section \ref{secSimulation}
provides simulation results. Section \ref{secConclusion} concludes
the paper.

{\it Notations}: We use bold faced characters to denote vectors. We
use $\bx[i]$ to denote the $i$th element of vector $\bx$. We use
${\mathbf{x}_{-i}}$ to denote the vector
$\left[\mathbf{x}[1],\cdots,\mathbf{x}[i-1],\mathbf{x}[i+1],\cdots
\mathbf{x}[N]\right]$. We use $[y,\mathbf{x}_{-i}]$ to denote a
vector $\mathbf{x}$ with its $i$th element replaced by $y$. We use
$\vee$ to denote componentwise maximization:
$\mathbf{x}\vee\mathbf{y}\triangleq\{\mathbf{z}|\mathbf{z}[i]=\max\{\mathbf{x}[i],
\mathbf{y}[i]\},\forall~i\}$. $\mathcal{N}\setminus i$ defines a
subset of $\mathcal{N}$: $\mathcal{N}\setminus i\triangleq\{j:
j\in\mathcal{N}, j\ne i\}$. The main notations used in this paper
are listed in Table \ref{tableSymbols}.

\vspace*{-0.1cm}
\begin{table*}[htb]
\caption{\small A List of Notations }\vspace*{-0.5cm}
\begin{center}
{\small
\begin{tabular}{|c |c | c|c|}
\hline
$\mathcal{N}$& The set of all users& $\mathcal{W}$& The set of all BSs\\
\hline $\mathcal{K}$& The set of all channels& $\mathcal{K}_w$& The
set of channels belongs to BS $w$\\
 \hline
$|h^k_i|^2$& The channel gain of user $i$ on channel $k$ &
$p^k_w$& The transmit power of BS $w$ on channel $k$\\
 \hline
 $\bar{p}_w$& The power budget for BS $w$ & $\beta^k_w$ & The user assignment of BS $w$ on channel $k$
\\
\hline $\mathbf{a}$& The association
profile& $\mathbf{a}_{-i}$ & The association profile without user $i$\\
 \hline
$r_i(\mathbf{a})$& The rate assigned to  user $i$&
$R_w(\mathbf{a})$& The optimal
throughput of BS $w$ under $\mathbf{a}$\\
\hline $R(\mathbf{a})$& The optimal system throughput under
$\mathbf{a}$ &$\mathcal{N}_w(\mathbf{a})$& The set of users
associated
to BS $w$\\
\hline
$U_i(\mathbf{a})$& The utility of user $i$& $T_i(\mathbf{a})$& The tax imposed on user $i$\\
 \hline
\end{tabular} } \label{tableSymbols}
\end{center}
\vspace*{-0.3cm}
\end{table*}

\vspace{-0.3cm}
\section{System Model and Problem Formulation}\label{secSystemModel}

We consider a service area with a set
$\mathcal{N}\triangleq\{1,2,\cdots,N\}$ of users served by a set
$\mathcal{W}\triangleq\{1,2,\cdots,W\}$ of BSs (or networks). Each
BS $w$ operates on the set of channels $\mathcal{K}_w$, with the
bandwidth of each channel equally set to be $\Delta f_w$. Let
$\mathcal{K}\triangleq\cup_{w\in\mathcal{W}}\mathcal{K}_w$ denote
the set of all channels. Suppose any two channels do not overlap,
i.e., $\mathcal{K}_w\cap\mathcal{K}_v=\emptyset,\ \forall\ w\ne v$.
Such assumption is justified for example in the multi-technology
HetNet or in the IEEE 802.22 cognitive radio Wireless Regional Area
Network (WRAN) \cite{fcc10}. In the latter network, a particular
geographical region may be served by multiple service providers
(SPs), or by multiple Access Points (APs) installed by a single SP.
When operating in the ``normal mode", the APs/SPs that serve the
same region indeed operate on non-overlapping portions of the
available spectrum, by using proper spectrum etiquette protocols
(see Section 6.22 in \cite{fcc10}).

Let $\left\{|h^k_{i}|^2\right\}_{k\in\mathcal{K}_w}$ denote the
channel gains of the set of channels from BS $w$ to user $i$; Let
$\left\{n^k_i\right\}_{k\in\mathcal{K}}$ denote the set of measured
noise powers at user $i$ on different channels. Both the channel
gains and the noise powers are considered as {\it private}
information to the users, as in the FDD mode they are measured at
the mobile devices and then fedback to the BSs.

Define a length $N$ vector $\mathbf{a}$ as  the {\it association
profile} in the network, with its $i$th element $\mathbf{a}[i]=w$
indicating that user $i$ is associated to BS $w$. Define
${\mathbf{a}_{-i}}\triangleq\left[\mathbf{a}[1],\cdots,\mathbf{a}[i-1],\mathbf{a}[i+1],\cdots
\mathbf{a}[N]\right]$ as an association profile in which user $i$
drops out of the network. For each BS $w$, denote the set of
associated users as
$\mathcal{N}_w(\mathbf{a})\triangleq\{i:\mathbf{a}[i]=w\}$, which is
a function of $\mathbf{a}$.

In a downlink OFDMA network, a BS $w\in\mathcal{W}$ can transmit to
a single user $i\in\mathcal{N}_w(\mathbf{a})$ on a given channel
$k\in\mathcal{K}_w$. Let
$\bfbeta_w\triangleq\{\beta^k_w\}_{k\in\mathcal{K}_w}$ be a feasible
channel assignment scheme for BS $w$, i.e.,
$\beta^k_w=i\in\mathcal{N}_w(\mathbf{a})$ means channel $k$ is
assigned to user $i$. Let
$\mathbf{p}_w\triangleq\{{p}^k_w\}_{{k\in\mathcal{K}_w}}$ be a
feasible power allocation scheme for BS $w$: $\mathbf{p}_w\ge
\mathbf{0}$, $\sum_{k\in\mathcal{K}_w}{p}^k_w\le \bar{p}_w$, where
$\bar{p}_w$ is the power budget for BS $w$. Let
$\bfbeta\triangleq\{\bfbeta_w\}_{w\in\mathcal{W}}$ and
$\mathbf{p}\triangleq\{\mathbf{p}_w\}_{w\in\mathcal{W}}$.

Let us define $r_i(\bfbeta, \mathbf{p}, \mathbf{a})$ as the
transmission rate that user $i$ can obtain under the resource
allocation scheme $(\bfbeta, \bp,\ba)$. With continuous rate
adaptation, this rate can be expressed as:{\small
\begin{align}
r_i(\bfbeta,
\mathbf{p},\mathbf{a})=\sum_{k\in\mathcal{K}_{\mathbf{a}[i]}}\Delta
f_{\mathbf{a}[i]}\log\left(1+\frac{|h^k_{i}|^2p^k_{\mathbf{a}[i]}}{\tau
n^k_i}\mathbf{1}\left\{\beta^k_{\mathbf{a}[i]}=i\right\}\right)\label{eqRate}
\end{align}}
where $\mathbf{1}\{\cdot\}$ is the indicator function; $\tau$ is the
capacity gap  which is determined by the target Bit Error Rate (BER)
as: $\tau=-\frac{\ln(5{\rm BER})}{1.5}$ (see \cite{goldsmith05}).

The objective of the resource allocation is to find the tuple
$(\bfbeta,\mathbf{p}, \mathbf{a})$ that achieves efficient spectrum
utilization within each BS/network while balancing the loads across
different BSs/networks. Mathematically, we formulate the overall
resource allocation problem as follows {\small
\begin{align}
\max_{\mathbf{a},\boldsymbol{\beta},\mathbf{p}}\quad&\sum_{w\in\mathcal{W}}\alpha_w
\sum_{i\in\mathcal{N}_w(\mathbf{a})}
r_i(\bfbeta,\mathbf{p},\mathbf{a})
\tag{\bf{SYS}}\nonumber \\
\textrm{s.t.}\quad&\mathbf{a}[i]\in\mathcal{W},~~\forall~i\in\mathcal{N},\nonumber\\
\quad& \beta^k_w\in\mathcal{N}_w(\mathbf{a}), \forall \  k \in\mathcal{K}_w, \forall \ w\in\mathcal{W},\nonumber\\
\quad& \mathbf{p}_w\ge \mathbf{0}, \
 \sum_{k\in\mathcal{K}_w}{p}^k_w\le \bar{p}_w, \
\forall~w\in\mathcal{W}\nonumber.
\end{align}}
The load balancing property of this formulation is manifested by: 1)
introducing the association as a decision variable for the users; 2)
including the weighting factors $\{\alpha_w\ge0\}_{w=1}^{W}$ in the
objective. The first factor enables the users to effectively avoid
congestion by switching to light-loaded BSs in a timely fashion,
while the second factor allows the network operator to further shift
the traffic to the BSs with larger weights. 

\vspace{-0.4cm}
\subsection{The BSs' Resource Management: Optimal Channel Assignment and Power
Allocation}\label{subBSProblem}

We first describe each BS's optimal resource management strategy.
Let us assume that each BS $w$ has perfect knowledge of downlink
{\it normalized} channel states of its associated users
$\mathbf{h}_w\triangleq\big\{\frac{h^k_{i}}{n^k_i}\big\}_{i\in\mathcal{N}_w(\mathbf{a}),k\in\mathcal{K}_w}$
(this assumption will be relaxed later).

First consider a simple case where a equal power allocation strategy
is used, that is: $~{p}^k_w=\frac{\bar{p}_w}{|\mathcal{K}_w|},~
~\forall~k\in\mathcal{K}_w$. Each BS $w$ then optimizes its
throughput by picking a suitable user to serve on each channel.
Mathematically, it solves the following
 channel assignment (CA) problem{\small
\begin{align}
\max_{{\boldsymbol{\beta}_w
}}\quad&\alpha_w\sum_{i\in\mathcal{N}_w(\mathbf{a})}r_i(\bfbeta,
\mathbf{p},\mathbf{a})\nonumber\tag{\bf{CA}}
\nonumber\\
\textrm{s.t.}\quad&
\beta^k_w\in\mathcal{N}_w(\mathbf{a}),\forall~k\in\mathcal{K}_w.\nonumber
\end{align}}
The optimal solution to this problem is to assign each channel to
the best user \cite{song05a}:{\small
\begin{align}
&\big(\beta_w^k\big)^*=i^*,~~\textrm{where}~i^*
\in\arg\max_{i\in\mathcal{N}_w(\mathbf{a})}\frac{|h^k_{i}|^2}{\tau
n^k_i}.\label{eqCA}
\end{align}}
On the other hand, if the BSs can optimize both its channel
assignment and power allocation, then a BS $w$ solves the following
channel assignment and power allocation (CAPA) problem:{\small
\begin{align}
\max_{{\boldsymbol{\beta}_w,\mathbf{p}_w }}\quad
&\alpha_w\sum_{i\in\mathcal{N}_w(\mathbf{a})}r_i(\bfbeta,\mathbf{p},\mathbf{a})\nonumber\tag{\bf{CAPA}}\\
\textrm{s.t.}\quad&\sum_{k\in\mathcal{K}_w}p^k_w\le
\bar{p}_w,~{p}^k_w\ge 0,~
\beta^k_w\in\mathcal{N}_w(\mathbf{a}),~~\forall~k\in\mathcal{K}_w.
\nonumber
\end{align}}
The optimal solution to this problem can be written in closed form
\cite{tse_m}:{\small
\begin{align}
\left\{ \begin{array}{l}
(\beta_w^k)^*=i^*,~~\textrm{where}~i^*\in\arg\max_{i\in\mathcal{N}_w(\mathbf{a})}\frac{|h^k_{i}|^2}{\tau n^k_i}\\
(p_w^k)^*=\left[{\lambda}-\frac{\tau n^k_{i^*}}{|h_{i^*}^k|^2}\right]^+ \\
\lambda\bigg(\sum_{k\in\mathcal{K}_w}\big(p^k_w\big)^*-\bar{p}_w\bigg)=0
\end{array}\right.\label{eqCAPA}
\end{align}}
where $\lambda\ge 0$ is the dual variable associated with the power
budget constraint. We note that the power constraint is binding at
the optimal solution:
$\sum_{k\in\mathcal{K}_w}({p}^k_w)^*=\bar{p}_w$.


With some abuse of notations, we use $r_i(\mathbf{a})$ to denote the
optimal rate for user $i$ obtained by using either the CA or CAPA
strategy (the actual strategy used will be indicated using a
superscript CA or CAPA when necessary). When $\mathbf{a}$ is fixed,
we denote the weighted optimal throughput of BS $w$ by: {$
R_w(\mathbf{a})\triangleq
\alpha_w\sum_{i\in\mathcal{N}_w(\mathbf{a})}r_i(\mathbf{a})$}.

\vspace{-0.3cm}
\subsection{The Overall Resource Management Problem: Complexity
Status}\label{subNP}

In this subsection, we investigate the complexity status of the
throughput optimization problem (SYS). 
A tuple $(\bfbeta^*, \mathbf{p}^*,\mathbf{a}^*)$ is an optimal
solution of the problem (SYS) only if each BS uses the CAPA
strategy. Although finding the CAPA solution is easy when the
user-BS association is fixed, the problem turns out to be
intractable when the association becomes an optimization variable.
The proof of this theorem is provided in the Appendix.
\newtheorem{T1}{Theorem}
\begin{T1}\label{theoremNP}
{\it Finding the optimal solution to the problem (SYS) is strongly
NP-hard. }
\end{T1}

When each BS is restricted to allocate resource by channel
assignment only, then the CA strategy is optimal for the per-BS
problem. In this case, the problem (SYS) is still intractable.

\newtheorem{C1}{Corollary}
\begin{C1}\label{corollaryNP}
{\it When the CA strategy is used for resource allocation within
each BS, finding the best BS association that maximizes the system
throughput is strongly NP-hard. }
\end{C1}

The above results establish the complexity status for the problem
(SYS) in a general network configuration. In a special case where
each BS operates on a {\it single} channel, the problem is
equivalent to a maximum weighted matching problem, which is
efficiently solvable.
\newtheorem{C2}{Corollary}
\begin{C1}\label{corollaryPoly}
{\it When $|\mathcal{K}_w|=1, \ \forall\ w\in\mathcal{W}$, problem
(SYS) is polynomial time solvable. }
\end{C1}

\vspace{-0.0cm}
\section{Mechanism Design for Joint BS
Association and Resource Allocation}\label{secGame}

The previous section analyzes the per-BS and the overall resource
allocation problem assuming complete information at the network
side. However, in an FDD network, there is an intrinsic {\it
asymmetry} in the available information at the BSs and at the users,
as the downlink channel and the noise powers are measured by the
users. Strategic (selfish) users can exploit such asymmetry of
information for their own benefit by tampering with the devices if
necessary \cite{Kavitha12}.  We now provide a simple illustration of
the potential inefficiency caused by the manipulation of channel
state information.

\newtheorem{E1}{Example}
\begin{E1}\label{example1}
Consider the network consisting of $1$ BS and $2$ users with $3$
channels. Let the noise power $n^k_i=1$ for all $i,k$, and let
$\tau=1$. Assume that the BS has a total power of $3$. The channel
gains are listed in Table \ref{tableChannelExample}.

\vspace*{-0.2cm}
\begin{table}[htb]
\caption{Channel Gains For Example 1.}
\begin{center}
\small{
\begin{tabular}{|c |c | c | c | c |}
\hline
 & $|h^1_{i}|^2$ & $|h^{2}_{i}|^2$ & $|h^{3}_{i}|^2$\\
 \hline
 \hline
 {\bf user 1 ($i=1$)} &{\bf 2} & {\bf 2} & 1 \\
 \hline
 {\bf user 2 ($i=2$)} &0.5 & 0.5& {\bf 2} \\
 \hline
\end{tabular} } \label{tableChannelExample}
\end{center}
\vspace*{-0.2cm}
\end{table}

When all the users report truthfully, and when the CA strategy is
used, user $1$ will be scheduled on channel $1$ and $2$, while user
$2$ will be scheduled on channel $3$. A  throughput of
$3\log(1+2)\approx 3.29$ ${\rm nats/s}$ can be obtained. When user
$1$ remains truthful but user $2$ becomes selfish, and it falsely
reports its channels as $(3,3,2)$ (instead of $(0.5,0.5,2)$), the BS
will assign {\it all} the channels to user $2$. After the channel
assignment, the {actual rate} that user $2$ obtains still depends on
its {\it true} channels \footnote{Such rate can be achieved via the
use of {\it rateless codes}. We refer the readers to \cite[Section
II]{Kavitha12} for detailed explanation of achieving such rate when
BSs do not have perfect knowledge of the actual channel.}. Thus a
throughput of $2\log(1+0.5)+\log(1+2)\approx 1.91$ ${\rm nats/s}$
will be obtained, which is only about $58\%$ of the optimal system
throughput. In contrast, user $2$'s untruthful behavior leads to its
own {\it rate increase} of over $70\%$, at the expense of starving
user $1$. \hfill$\blacksquare$
\end{E1}

\subsection{The VCG mechanism}\label{subVCG}

The optimization of system performance when strategic users have
private information can be formulated as a {\em mechanism design}
problem. Assuming users have quasilinear utility, a system of
incentives (interference taxes) may be put  in place in order to
align individual users' preferences with the goal of optimizing
system performance. The goal therefore is to find the interference
taxes that support the implementation of efficient resource
allocation in dominant strategies, i.e. for each user, the truthful
revelation of channel state information is optimal regardless of the
information reported by all other users. The search for mechanisms
is typically restricted to the class of {\em direct} mechanisms in
which users report their private information to a third-party, which
in turn allocates resources and implements a system of incentives
via taxes.

The celebrated VCG mechanism achieves this goal by having users
report their  privately held information on channel states to a
central controller (CU), who computes the {\it globally optimal}
solution of (SYS) {given} the {\it reported} information. The CU
then assigns the users to the BSs and a given rate according to the
solution of the optimal solution of (SYS). Each user, when
attempting to manipulate the allocation of resources by misreporting
channel state information, is penalizes for the deterioration of
system performance for all other users. This is the basis for the
VCG mechanism being strategy-proof, i.e. for each individual user,
truthful revelation of channel state information is optimal
regardless of the information reported by all other users. It should
be also emphasized here that any other direct and strategy-proof
mechanism implementing the solution to (SYS) is an instance of the
VCG mechanism with interference taxes modified by a constant (see
\cite[Corollary 5.1]{borgers10}). Unfortunately, in the previous
section we showed that finding the the global optimal solution of
(SYS) is an NP-hard problem. Thus, a computationally tractable
direct mechanism cannot be both strategy-proof and efficient.

Our strategy for designing a computationally tractable mechanism is
to relax the requirement of optimality so that an {\em
approximately} optimal solution of (SYS) can be implemented in
dominant strategies. In the mechanism proposed below, tractability
is achieved by: {\em (i)} decentralizing the resource allocation
decisions to each BS and implementing the VCG mechanism in a {\it
per BS} basis (section \ref{subIC} below); {\em (ii)} allowing the users to dynamically adjust
their choices of association (section \ref{subuserProblem} below).

\vspace{-0.2cm}
\subsection{Implementing VCG at each BS with Fixed Association}\label{subIC}
We formally describe the implementation of the VCG mechanism for {\em given} user-BS association profile $\mathbf{a}$.
Recall that optimal per-BS strategies were described in Section \ref{subBSProblem}.

Define the
normalized channels as {\small $
\mathbf{h}_{i,w}\triangleq\big\{\frac{h^k_{i}}{n^k_{i}}\big\}_{k\in\mathcal{K}_w};
\mathbf{h}_{-i,w}\triangleq\{\mathbf{h}_{j,w}\}_{j\in\mathcal{N}_w(\mathbf{a})\setminus
i}.\nonumber $} Let $\mathbf{h}_w\triangleq[\mathbf{h}_{i,w},
\mathbf{h}_{-i,w}]$. Define user $i$'s {\it reported} normalized
channels as $\widehat{\mathbf{h}}_{i,w}$. Define
$\widehat{\mathbf{h}}_{-i,w}$ and $\widehat{\mathbf{h}}_w$
similarly. When we take untruthfulness into consideration, a user
$i$'s rate depends on the following two terms: 1) the {reported}
normalized channel, denoted as $\widehat{\mathbf{h}}_w$, by which
the BS makes the resource allocation decision; 2) the {actual}
normalized channel $\mathbf{h}_{i,w}$, by which user $i$ experiences
the {\it actual} rate.  We signify such dependencies by using
$r_{i}(\mathbf{a}; \mathbf{h}_{i,w}, \widehat{\mathbf{h}}_w)$ to
denote user $i$'s rate. If the information reported by the users is
$\widehat{\mathbf{h}}_{w}$, a tax $T_i(\mathbf{a};
\widehat{\mathbf{h}}_{w})$ will be levied upon user $i$, and its net
utility is{\small
\begin{align}
U_i(\mathbf{a}; {\mathbf{h}}_{i,w},
\widehat{\mathbf{h}}_{-w})=\alpha_w r_i(\mathbf{a};\mathbf{h}_{i,w}
\widehat{\mathbf{h}}_{w})-T_i(\mathbf{a};\widehat{\mathbf{h}}_{w}).\label{eqUtilityIC}
\end{align}}
The tax assessed on user $i$ is computed based on the {\it reported}
channels. It is given as the {\it total rate improvement} that the
set of remaining users $\mathcal{N}_{w}(\mathbf{a})\setminus i$ can
obtain {\it if user $i$ leaves BS $w$:}{\small
\begin{align}
 T_i(\mathbf{a}; \widehat{\mathbf{h}}_{w})\triangleq \underbrace{\sum_{ j\in\mathcal{N}_{w}(\mathbf{a}_{-i})}
\alpha_w r_j(\mathbf{a}_{-i}; \widehat{\mathbf{h}}_{j, w},
\widehat{\mathbf{h}}_{-i,w})}_{\rm optimal~throughput~when~i~
drops~out}-\underbrace{\sum_{
j\in\mathcal{N}_{w}(\mathbf{a})\setminus i}\alpha_w r_j(\mathbf{a};
\widehat{\mathbf{h}}_{j,w}, \widehat{\mathbf{h}}_{w})}_{\rm
optimal~throughput~of~other~users~when~i~is~present}.\label{eqTax}
\end{align}}
The tax expressed in
\eqref{eqTax} ensures that each user has an incentives to act
truthfully. To see why this is note that when all the users truthfully report their channels, BS
$w$ can maximize its throughput. For example, the optimal throughput
is better than that achieved when user $i$ reports
untruthfully:
{\small
\begin{align}
r_i({\mathbf{a}}; {\mathbf{h}}_{i,w},
[\widehat{\mathbf{h}}_{i,w},{\mathbf{h}}_{-i,w}])+\sum_{j\in\mathcal{N}_w(\mathbf{a})\setminus
i}r_j({\mathbf{a}}; {\mathbf{h}}_{j,w},
[\widehat{\mathbf{h}}_{i,w},{\mathbf{h}}_{-i,w}])\le
\sum_{j\in\mathcal{N}_w(\mathbf{a})}r_j({\mathbf{a}};
{\mathbf{h}}_{j,w}, \mathbf{h}_{w})\label{eqRateMaximization}
\end{align}}

From this property we can derive the following key inequality{\small
\begin{align}
&r_i({\mathbf{a}}; {\mathbf{h}}_{i,w},
\widehat{\mathbf{h}}_{w})+\hspace{-0.2cm}\sum_{
j\in\mathcal{N}_{w}(\mathbf{a})\setminus i}r_j(\mathbf{a};
\widehat{\mathbf{h}}_{j,w}, \widehat{\mathbf{h}}_{w}) \le
{r_i({\mathbf{a}}; {\mathbf{h}}_{i,w},
[\mathbf{h}_{i,w},\widehat{\mathbf{h}}_{-i,w}])+\hspace{-0.2cm}\sum_{
j\in\mathcal{N}_{w}(\mathbf{a})\setminus i}r_j(\mathbf{a};
\widehat{\mathbf{h}}_{j,w},
[\mathbf{h}_{i,w},\widehat{\mathbf{h}}_{-i,w}])}\label{IC}.
\end{align}}
The sum of the two terms in the right hand side of \eqref{IC}
represents the sum rate achieved by all the users in BS $w$ {\it
assuming the actual channels} are
{\small$[\mathbf{h}_{i,w},\widehat{\mathbf{h}}_{-i,w}]$}, and the
users also truthfully report
{\small$[\mathbf{h}_{i,w},\widehat{\mathbf{h}}_{-i,w}]$}.
Consequently, this inequality follows from the fact that the BS's
resource management strategy maximizes the sum rate when the
channels are truthfully reported.

Now suppose user $i$ reports untruthfully (i.e.,
$\mathbf{h}_{i,w}\ne\widehat{\mathbf{h}}_{i,w}$), then we have
{\small
\begin{align}
&U_i(\mathbf{a};{\mathbf{h}}_{i,w},
\widehat{\mathbf{h}}_{w})\nonumber\\
&\stackrel{(a)}=r_i(\mathbf{a}; {\mathbf{h}}_{i,w},
\widehat{\mathbf{h}}_{w})-\bigg(\sum_{
j\in\mathcal{N}_{w}(\mathbf{a}_{-i})} r_j(\mathbf{a}_{-i};
\widehat{\mathbf{h}}_{j, w}, \widehat{\mathbf{h}}_{-i,w})-\sum_{
j\in\mathcal{N}_{w}(\mathbf{a})\setminus i}r_j(\mathbf{a};
\widehat{\mathbf{h}}_{j,w}, \widehat{\mathbf{h}}_{w})\bigg)\nonumber\\
 &\stackrel{(b)}\le {r_i({\mathbf{a}}; {\mathbf{h}}_{i,w},
[\mathbf{h}_{i,w},\widehat{\mathbf{h}}_{-i,w}])+\sum_{
j\in\mathcal{N}_{w}(\mathbf{a})\setminus i}r_j(\mathbf{a};
\widehat{\mathbf{h}}_{j,w},
[\mathbf{h}_{i,w},\widehat{\mathbf{h}}_{-i,w}])}-\sum_{
j\in\mathcal{N}_{w}(\mathbf{a}_{-i})} r_j(\mathbf{a}_{-i};
\widehat{\mathbf{h}}_{j, w},
\widehat{\mathbf{h}}_{-i,w})\nonumber\\
&\stackrel{(c)}=U_i(\mathbf{a};{\mathbf{h}}_{i,w},
[\mathbf{h}_{i,w},\widehat{\mathbf{h}}_{-i,w}])\nonumber
\end{align}}
where $(a)$ and $(c)$ are from the definitions of the utility in
\eqref{eqUtilityIC}; $(b)$ is from \eqref{IC}. We have then
established desired result.

In the reminder of this paper, we will assume that the VCG mechanism is implemented at each BS.
Thus, we will simply write {\small$r_i( \mathbf{a})$} instead
of {\small$r_i( \mathbf{a}; \mathbf{h}_{i,w},
\widehat{\mathbf{h}}_{w})$}. The user $i$'s tax term \eqref{eqTax}
and utility term \eqref{eqUtilityIC} can be simplified as (assuming
$\mathbf{a}[i]=w$){\small
\begin{align}
 T_i(\mathbf{a})&\triangleq \sum_{ j\in\mathcal{N}_{w}(\mathbf{a}_{-i})}
\alpha_w  r_j(\mathbf{a}_{-i})-\sum_{
j\in\mathcal{N}_{w}(\mathbf{a})\setminus i} \alpha_w  r_j(\mathbf{a})\label{eqTaxSimple}\\
U_i(\mathbf{a})&\triangleq \alpha_w
 r_i(\mathbf{a})-T_i(\mathbf{a})=\sum_{
j\in\mathcal{N}_{w}(\mathbf{a})}\alpha_w  r_j(\mathbf{a})-\sum_{
j\in\mathcal{N}_{w}(\mathbf{a}_{-i})} \alpha_w
 r_j(\mathbf{a}_{-i})\label{eqUtilitySimple}.
\end{align}}
In summary, by using the VCG mechanism within each BS, all the users
will act truthfully, which in turn allows the BSs to optimally
implement their resource allocation strategies. It is important to
note here, that even in the ideal scenario where all the users
behave truthfully, the tax and utility function defined in
\eqref{eqTaxSimple} and \eqref{eqUtilitySimple} are still extremely
useful. As will be seen in the subsequent sections, they lead to
simple and efficient network-wide resource allocation.

\vspace{-0.4cm}
\subsection{The User-BS Association Game}
\label{subuserProblem}

Suppose users are allowed to autonomously select which BS to connect to.
Assuming each BS implements a
VCG mechanism, we are left with a user-BS association game.
We will occasionally use the superscripts ${\rm CA}$ or ${\rm CAPA}$ to
specify the strategies used by the BSs. Let us define a
non-cooperative BS association game as: $
\mathcal{G}\triangleq\{\mathcal{N}, \{\chi_i\}_{i\in\mathcal{N}},
\{U_i(\cdot)\}_{i\in\mathcal{N}}\}$, where $\chi_i=\mathcal{W}$ is
the strategy space of user $i$; $U_i(\cdot)$ is the utility of user
$i$ as defined in \eqref{eqUtilitySimple}.

Interestingly, unlike most conventional games, in game
$\mathcal{G}$, the interdependencies of the users' strategies are
only {\it implicitly} given. For example, suppose $\mathbf{a}[i]=w$,
$\mathbf{a}[j]=q$. In order to assess the impact of user $i$'s
change of association from BS $w$ to BS $q$ on user $j$'s utility,
BS $q$'s resource allocation problem (either CA or CAPA) needs to be
solved. There is no closed-form expression governing the users'
interdependencies. This unique property of the game makes our
subsequent analysis, particularly the efficiency of the NE of game
$\mathcal{G}$, very involved.

 We first characterize the utility function
$U_i(\mathbf{a})$ and the tax function $T_i(\mathbf{a})$.
\newtheorem{P1}{Proposition}
\begin{P1}\label{propUE}
{\it  When all the BSs use either the CA or the CAPA strategy,
$T_i(\mathbf{a})\ge 0,~\forall~i\in\mathcal{N}$. Moreover, the
users' utility functions are bounded: $0\le U_i(\mathbf{a})\le
\alpha_{\mathbf{a}[i]}r_i(\mathbf{a})$.}
\end{P1}
{\it Proof:}~For notational simplicity, let $w=\mathbf{a}[i]$ be the
association of user $i$. Suppose that the BSs use the CA strategy.
Observe that as each channel vacated by user $i$'s departure will be
re-assigned to some user $j\in\mathcal{N}_w(\mathbf{a})\setminus i$,
and the assignment of all the other channels remains the same.
Consequently, the rates of those users that have new channels will
increase. This leads to $T^{\rm CA}_i(\mathbf{a})\ge 0$, which in
turn implies that $U^{\rm CA}_i(\mathbf{a})\le \alpha_{w} r^{\rm
CA}_i(\mathbf{a})$. To show $U^{\rm CA}_i(\mathbf{a})\ge 0$, observe
{\small
\begin{align}
\max_{j\in\mathcal{N}_w(\mathbf{a})\setminus
i}\frac{|h^k_{j}|^2}{n^k_j}\le
\max_{j\in\mathcal{N}_w(\mathbf{a})}\frac{|h^k_{j}|^2}{n^k_j}, \
\forall~k\in\mathcal{K}_w.\label{eqChannelBetterCA}
\end{align}}
This inequality combined with the monotonicity of the log function
and the structure of the solution for the CA problem (cf.
\eqref{eqCA}) implies:{\small
\begin{align}
\sum_{j\in\mathcal{N}_w(\mathbf{a}_{-i})}\alpha_w r^{\rm
CA}_j(\mathbf{a}_{-i})\le
\sum_{j\in\mathcal{N}_w(\mathbf{a})\setminus i}\alpha_w r^{\rm
CA}_j(\mathbf{a})+\alpha_w r^{\rm CA}_i(\mathbf{a}).
\end{align}}
 Rearranging terms, and plugging in the definition of tax in \eqref{eqTaxSimple}, we have: $ \alpha_w r^{\rm
CA}_i(\mathbf{a})-T^{\rm CA}_i(\mathbf{a})=U^{\rm
CA}_i(\mathbf{a})\ge 0$. The CAPA case can be argued along the same
lines.\hfill$\blacksquare$

\vspace{-0.4cm}
\subsection{Characterization of Nash Equilibria}\label{secPropoertyNE}
In this subsection we present a series of results characterizing the
pure NEs for game $\mathcal{G}$.

For a fixed $\mathbf{a}$, we define $BR_i(\mathbf{a})$ as the set of
``better-reply" BSs for user $i$:{\small
\begin{align}
BR_i(\mathbf{a})\triangleq\big\{w|U_i([w,\mathbf{a}_{-i}])>U_i([\mathbf{a}[i],\mathbf{a}_{-i}]),
w\in\mathcal{W}\big\}.\label{eqBRSet}
\end{align}}
The pure strategy NE of the game $\mathcal{G}$ is a profile
$\mathbf{a}^*$ in which
$BR_i(\mathbf{a}^*)=\emptyset,~\forall~i\in\mathcal{N}$.
Equivalently, all users prefer to stay in their current BSs: {\small
$ U_i\left(\mathbf{a}^*\right)\ge \max_{w\in\mathcal{W}}U_i\left([w,
\mathbf{a}^*_{-i}]\right),~~\forall~i\in\mathcal{N}\label{eqNES} .
$}

Let $R(\mathbf{a})\triangleq\sum_{w\in\mathcal{W}}\alpha_w
R_w(\mathbf{a})$ denote the weighted system throughput for fixed
association $\mathbf{a}$. Our first result analyzes the existence of
the pure NE of game $\mathcal{G}$. The proof can be found in the
Appendix.
\newtheorem{T2}{Theorem}
\begin{T1}\label{theoremExistenceS}
{\it The game $\mathcal{G}$ must admit at least one pure NE. In
particular, the association profile
$\widetilde{\mathbf{a}}\in\arg\max_{\mathbf{a}}R(\mathbf{a})$ must
be a pure NE of this game.}
\end{T1}

The existence of pure NE for the game $\mathcal{G}$ could be
attributed to the tax charged by the BSs. Without such tax, there
could be no pure NE. To illustrate, define a new game in which users
are not charged with taxation, and their utilities are just their
rates: $\widetilde{\mathcal{G}}\triangleq\{\mathcal{N},
\{\chi_i\}_{i\in\mathcal{N}}, \{r_i(\cdot)\}_{i\in\mathcal{N}}\}$.
We claim that if all the BSs use either CA or CAPA strategy, this
game does not always admit a pure NE. We show this claim by giving
two counterexamples.
\newtheorem{E2}{Example}
\begin{E1}\label{example2}
When the BSs use CA strategy, consider a network with $W=2$, $N=3$,
$\alpha_w=1$ and $|\mathcal{K}_w|=2$, $\forall \ w$. The channel
gains are given in the top part of Table \ref{tableChannel1}. Let
$n^k_i=1,~\forall~i,k$, $\bar{p}_w=2,~\forall~w$. When BSs use CAPA
strategy, consider a network with $W=2$, $N=3$, $\alpha_w=1$ and
$|\mathcal{K}_w|=2$, $\forall \ w$. The channel gains are given in
Table \ref{tableChannel2}. Let $n^k_i=1,~\forall~i,k$,
$\bar{p}_w=5,~\forall~w$. For both examples, we show in Table
\ref{tableBestReply} that in every possible association profile,
there exists at least one user whose better-reply set is nonempty.
\hfill $\blacksquare$
\end{E1}

\begin{table}[ht]
\begin{minipage}[b]{0.5\linewidth}
\caption{Channel Gains For Example \ref{example2}--CA Case.}
\vspace{-0.2cm} \centering \small{
\begin{tabular}{|c |c | c | c | c |}
\hline
 CA case & $|h^{1}_{i}|^2$ & $|h^{2}_{i}|^2$ & $|h^{3}_{i}|^2$ & $|h^{4}_{i}|^2$ \\
 \hline
 i=1 &2 & 0.1 & 2.2 &  0.1\\
 \hline
 i=2 &0.5 & 2.5& 0.1 & 2.6 \\
 \hline
 i=3 &0.1 & 2.4 & 2.3 &  0.2\\
 \hline
\end{tabular} } \label{tableChannel1}
\vspace*{-0.3cm}
\end{minipage}
\hspace{0.3cm}
\begin{minipage}[b]{0.5\linewidth}
\caption{Channel Gains For Example \ref{example2}--CAPA Case.}
\vspace{-0.2cm} \centering \small{
\begin{tabular}{|c |c | c | c | c |}
\hline
 CAPA case & $|h^{1}_{i}|^2$ & $|h^{3}_{i}|^2$ & $|h^{3}_{i}|^2$ & $|h^{4}_{i}|^2$ \\
 \hline
 i=1 &$\frac{1}{5}$ & $\frac{1}{5}$ & $\frac{1}{6.4}$ &  $\frac{1}{11}$\\
 \hline
 i=2 &$\frac{1}{6}$ & 0& 0 & $\frac{1}{8}$ \\
 \hline
 i=3 &0 & $\frac{1}{4}$ & $\frac{1}{6}$ &  0\\
 \hline
\end{tabular} } \label{tableChannel2}
\vspace*{-0.1cm}
\end{minipage}
\end{table}

\begin{table}[htb]
\vspace*{-0.1cm} \caption{The Better-Reply Sets for uses Under
Different System Association Profiles.}\vspace*{-0.3cm}
\begin{center}
\small{
\begin{tabular}{|c |c| c|} \hline
 {\bf  Association Profile} & {\bf Better-Reply Set (CA Example)} & {\bf Better-Reply Set (CAPA Example)}\\
 \hline
 $[1,1,1]$& $BR^{CA}_3([1,1,1])=2$ &$BR^{CAPA}_2([1,1,1])=2$ \\
  \hline
 $[1,1,2]$& $BR^{CA}_2([1,1,2])=2$ &$BR^{CAPA}_2([1,1,2])=2$\\
  \hline
 $[1,2,2]$& $BR^{CA}_3([1,2,2])=1$ &$BR^{CAPA}_3([1,2,2])=1$\\
  \hline
 $[1,2,1]$& $BR^{CA}_1([1,2,1])=2$ &$BR^{CAPA}_1([1,2,1])=2$\\
  \hline
 $[2,2,1]$& $BR^{CA}_2([2,2,1])=1$ &$BR^{CAPA}_2([2,2,1])=1$\\
  \hline
 $[2,1,1]$& $BR^{CA}_3([2,1,1])=2$ &$BR^{CAPA}_3([2,1,1])=2$\\
  \hline
 $[2,1,2]$& $BR^{CA}_1([2,1,2])=1$ &$BR^{CAPA}_1([2,1,2])=1$\\
  \hline
 $[2,2,2]$& $BR^{CA}_1([2,2,2])=1$ &$BR^{CAPA}_1([2,2,2])=1$\\
 \hline
\end{tabular} } \label{tableBestReply}
\end{center}
\vspace*{-0.2cm}
\end{table}

Example \ref{example2} illustrates that it is the interference tax
imposed by the BSs that ensures the existence of the pure NE for
game $\mathcal{G}$. In fact, such tax also guarantees the efficiency
of the outcome of the game. Theorem \ref{theoremExistenceS} asserts
that the {\it maximum} weighted throughput achievable by all the NEs
is the same as the optimal system weighted throughput. In the
following, we further provide a lower bound for the efficiency of
the NEs. Central to the derivation of such lower bound is certain
submodular property of the per-BS throughput function $R_w(\cdot)$.
Note that $R_w(\mathbf{a})$ depends on the association profile
$\mathbf{a}$ only through the set of associated users
$\mathcal{N}_w(\mathbf{a})$. We can then rewrite $R_w(\mathbf{a})$
as $R_w(\mathcal{N}_w(\mathbf{a}))$, which is expressed as a
function of the set of associated users. Then we say that
$R_w(\cdot)$ is {\it submodular} if the following is true{
\begin{align}
&R_w(\mathcal{G}\cup\{i\})-R_w(\mathcal{G})\le
R_w(\mathcal{M}\cup\{i\})-R_w(\mathcal{M}),~\forall~i\in\mathcal{N},\
\forall~\mathcal{M}\subseteq\mathcal{G}\subseteq\mathcal{N}.\label{eqDefSubmodularity2}
\end{align}}
The submodularity implies that there is a marginal decrease of
throughput when the total number of associated users increases. In
\cite{tse98part1}, Tse and Hanly have shown that for a fixed power
allocation policy without the total power constraint, the capacity
of a fading multiple access channel is a submodular function.
However, in our case showing the submodularity of the throughput
$R_w(\cdot)$ is much more involved, as our resource allocation is
the solution to the underlying optimization problems, hence it is
{dynamic} with respect to the set of associated users.

Once the submodularity property is shown, we can utilize a result
from Vetta \cite{vetta02} to obtain the desired lower bound. In
particular, reference \cite{vetta02} introduces the notion of {\it
valid-utility games}, for which lower bounds for the efficiency of
the NE is $\frac{1}{2}$. We will show that our BS selection game
$\mathcal{G}$ belongs to the family of valid-utility games.

\newtheorem{T3}{Theorem}
\begin{T1}\label{theoremAppRatio}
{\it The weighted system throughput achieved in any NE of the game
$\mathcal{G}$ must be at least half of that achieved under the
optimal user-BS assignment.}
\end{T1}
{\it Proof:} It is easy to check that $R_w(\cdot)$ has a {\it
monotonicity} property: {\small $R_w(\mathcal{M})\le
R_w(\mathcal{G}),~\forall~\mathcal{M}\subseteq\mathcal{G}$}. We then
claim that $R_w(\cdot)$ satisfies \eqref{eqDefSubmodularity2}. We
only give proof for the (more difficult) CAPA case, the CA case is a
straightforward extension. For simplicity of notations, we let BS
$w$ operate on all channels $\mathcal{K}$, set $n^k_j=1$ for all
$j,k$, and let $\alpha_w=1$.

Fix two sets $\mathcal{M},\mathcal{G}$ with
$\mathcal{M}\subseteq\mathcal{G}$, fix an arbitrary user $i$ with
arbitrary channel gains. Define three vectors
$\mathbf{g},\mathbf{m},\mathbf{h}\in\mathbb{R}^{K}_{+}$, with their
elements given as{\small
\begin{align}
\mathbf{g}[k]=\max_{j\in\mathcal{G}}|h^k_{j}|^2,~\mathbf{m}[k]=\max_{j\in\mathcal{M}}|h^k_{j}|^2,~
\mathbf{h}[k]=|h^k_{i}|^2.
\end{align}}
Note $\mathbf{g}$ and $\mathbf{m}$ represent the {\it best channel
gain} on each channel for the set of users $\mathcal{G}$ and
$\mathcal{M}$, respectively. From the fact that
$\mathcal{M}\subseteq\mathcal{G}$, we have that $\mathbf{m}\le
\mathbf{g}$. Note that the throughput obtained by BS $w$ using the
CAPA strategy is dependent on the set of associated users only
through the best channel vector. As a result, we can also express
$R_w(\mathcal{G})$ as $R_w(\mathbf{g})$, and $R_w(\mathcal{G}\cup
\{i\})$ as $R_w(\mathbf{g}\vee\mathbf{h})$. In this notation, the
submodular property \eqref{eqDefSubmodularity2} is equivalent to
{\small
\begin{align}
R_w(\mathbf{g}\vee\mathbf{h})-R_w(\mathbf{g})\le
R(\mathbf{m}\vee\mathbf{h})-R_w(\mathbf{m}),~\forall~\mathbf{h}\ge
0,~\mathbf{g}\ge\mathbf{m}\ge 0 . \label{eqDefSubmodularity3}
\end{align}}
In the same token, the monotonicity of $R_w(\cdot)$ can be expressed
as{\small
\begin{align}
R_w(\mathbf{g})\ge
R_w(\mathbf{m}),\forall~\mathbf{g}\ge\mathbf{m}\ge 0.
\end{align}}
We then present a sufficient condition for
\eqref{eqDefSubmodularity3} which is easier to verify. Let
$\mathbf{e}_k$ be a $K\times 1$ unit vector with its $k^{th}$
element being $1$.  Write
$\mathbf{h}=\sum_{k=1}^{K}\mathbf{e}_k\mathbf{h}[k]$. Then we have
{\small
\begin{align}
&R_w(\mathbf{g}\vee\mathbf{h})-R_w(\mathbf{g})=\big[R_w(\mathbf{g}\vee\sum_{k=1}^{K}\mathbf{e}_k
\mathbf{h}[k])-R_w(\mathbf{g}\vee\sum_{k=1}^{K-1}\mathbf{e}_k
\mathbf{h}[k])\big]+\cdots +\big[R_w(\mathbf{g}\vee\mathbf{e}_1
\mathbf{h}[1])-R_w(\mathbf{g})\big]\nonumber\\
&R_w(\mathbf{m}\vee\mathbf{h})-R_w(\mathbf{m})=\big[R_w(\mathbf{m}\vee\sum_{k=1}^{K}\mathbf{e}_k
\mathbf{h}[k])-R_w(\mathbf{m}\vee\sum_{k=1}^{K-1}\mathbf{e}_k
\mathbf{h}[k])\big]+\cdots +\big[R_w(\mathbf{m}\vee\mathbf{e}_1
\mathbf{h}[1])-R_w(\mathbf{m})\big].\nonumber
\end{align}}
In order for \eqref{eqDefSubmodularity3} to be true, it is
sufficient that for all $k\in\mathcal{K}$, the following is true
{\small
\begin{align}
R_w(\mathbf{g}\vee\mathbf{e}_k\mathbf{h}[k])-R_w(\mathbf{g})\le
R_w(\mathbf{m}\vee\mathbf{e}_k\mathbf{h}[k])-R_w(\mathbf{m}),~\forall~\mathbf{h}\ge
0, ~ \mathbf{g}\ge\mathbf{m}\ge 0. \label{eqDefSubmodularity4}
\end{align}}
Condition \eqref{eqDefSubmodularity4} allows us to verify the
submodular condition on a {\it channel by channel} basis. Partition
the set $\mathcal{K}$ into two sets: $\mathcal{Q}=\{k|
\mathbf{m}[k]=\mathbf{g}[k]\}$, $\overline{\mathcal{Q}}=\{k|
\mathbf{m}[k]<\mathbf{g}[k]\}$. We can show that
\eqref{eqDefSubmodularity4} is true for all $k\in\mathcal{Q}$ and
$k\in\overline{\mathcal{Q}}$. The proof for this result is given in
the Appendix.

To this point we have shown that $R_w(\cdot)$ is submodular and
monotone. From Proposition \ref{propUE} we have that
$\sum_{w}\sum_{i\in\mathcal{N}_{w}(\mathbf{a})}U_i(\mathbf{a})\le
\sum_w R_w(\mathbf{a})$. Additionally, the definition of
$U_i(\cdot)$ ensures that it is equal to the difference of the
system throughput {\it with and without} user $i$ (cf.
\eqref{eqUtilitySimple}). As a result, game $\mathcal{G}$ is a valid
utility game, and we can apply \cite[Theorem 3]{vetta02} to deduce
that any NE of the game achieves at least a half of the optimal
weighted throughput. This completes the proof.\hfill$\blacksquare$

We emphasize that all the results derived in Section
\ref{subuserProblem} and \ref{secPropoertyNE} hold true {\it
regardless} of the presence of the untruthful users, as long as the
BSs implement the taxation for each user as specified in
\eqref{eqTaxSimple} and \eqref{eqUtilitySimple}. This is because the
association game $\mathcal{G}$ is built upon the assumption that the
BSs use the VCG mechanism, and that the users are always truthful.

\vspace{-0.3cm}
\section{A Dynamic Mechanism}\label{secAlgorithm}
In this section we introduce a mechanism that allows the users and
the BSs to jointly compute a NE of the game $\mathcal{G}$, which is
a high quality solution for the joint BS selection and resource
allocation problem. All the results in this section are applicable
to both games $\mathcal{G}^{\rm CA}$ and $\mathcal{G}^{\rm CAPA}$.
Suppose each user maintains a length $M$ memory that operates in a
first in first out fashion. Each user's memory is used to store its
best associations in the last $M$ iterations.

We first briefly describe the main steps of the proposed mechanism.
It alternates between a BS optimization step and a user optimization
step. When it is the BSs' turn to act, based on the current set of
associated users, each of the BSs optimally allocates the resources
in its own cell using the VCG mechanism (cf. Section
\ref{subBSProblem} and \ref{subIC}). From the system perspective, in
this step, the tuple $(\bfbeta,\mathbf{p})$ is updated while holding
the association $\mathbf{a}$ fixed. When it is the users' turn to
act, each of them first computes its current best BS (in terms of
achieved individual utility) according to the current association
profile. It then pushes the best BS into its memory, and {randomly
samples} one BS from its memory for actual association. As we will
see later, in this step, $\mathbf{a}$ is updated while fixing
$(\bfbeta,\mathbf{p})$. The sampling step is the key to establish
the convergence of the proposed mechanism. The proposed mechanism is
detailed in Table \ref{tableAlgorithm}, where the superscript $(t)$
denotes the iteration number.

\begin{table}[htb]
\begin{center}
\vspace{-0.15cm} \caption{The Proposed Mechanism}\vspace*{-0.1cm}
\label{tableAlgorithm} {\small
\begin{tabular}{|l|}
\hline
S1) {\bf Initialization}: Let t=0, let the users choose their nearest BSs.\\
S2) {\bf BS Optimization}: Based on current
 $\mathbf{a}^{(t)}$, each BS implements a VCG mechanism.\\
S3) {\bf User Optimization}: For each user $i\in\mathcal{N}$  \\
\quad S3-1) {\bf Compute the Best BS}: Compute
$BR_i(\mathbf{a}^{(t)})$; If $BR_i(\mathbf{a}^{(t)})\ne\emptyset$,\\
\quad randomly select $w^{*(t)}_i\in
BR_i(\mathbf{a}^{(t)})$; otherwise, set $w^{*(t)}_i=\mathbf{a}^{(t)}[i]$ \\
\quad S3-2) {\bf Update Memory}: Shift $w^{*(t)}_i$ into the front
of the
memory; \\
\quad if $t\ge M$, shift $w^{*(t-M)}_i$ out from the end of the memory \\
\quad S3-3) {\bf Determine the Next BS Association}: Uniformly sample the user $i$'s memory;\\
\quad   obtain a BS index as $\mathbf{a}^{(t+1)}[i]$\\
S4) {\bf Continue}: If $\mathbf{a}^{(t+1)}=\mathbf{a}^{(t+1-m)}$ for
$m=1,\cdots,M$, stop.\\
\quad Otherwise, let t=t+1, go to S2).\\
 \hline
\end{tabular}}
\end{center}
\vspace*{-0.5cm}
\end{table}

One benefit of the proposed mechanism is that it allows the users to
update at the same time {\it without} explicit coordination of their
update sequences. A possible alternative is a sequential
implementation that allows a {\it single} user to update in each
iteration. However such scheme is undesirable as it requires
significant coordination efforts among all the users/BSs. 

An important feature of the mechanism is that each of its steps can
be implemented in a distributed fashion. The following two assumptions on the
network are needed for such purpose:
\begin{enumerate}
\item {\it Local} channel information is known by each BS. That is, each
BS $w$ has the knowledge of
$\left\{|h^k_{i}|^2\right\}_{k\in\mathcal{K}_w, i\in\mathcal{N}}$,
but not the channels related to other BSs. Note that in an FDD
system, this information is obtained via user feedback,  the
truthfulness of which is ensured by implementing the per-BS VCG
mechanism;

\item Each BS has a feedback channel to all the potential
users.
\end{enumerate}
Under the above assumptions, the mechanism can be implemented
distributedly. In the BSs' optimization step, the BSs compute the
taxes and perform their per-cell resource allocation (cf. Section
\ref{subBSProblem} and \ref{subIC}). They are not required to have
the knowledge of the operational conditions or channel states
related to other BSs. In the users' optimization step, to compute
the set $BR_i(\ba^{(t)})$, each user $i$ needs to know
$U_i([w,\ba^{(t)}_{-i}]),~\forall~w$ (cf. \eqref{eqBRSet}). Each of
these quantities can be expressed as{\small
\begin{align}
U_i([w,\ba^{(t)}_{-i}])=\alpha_w
r_i([w,\ba^{(t)}_{-i}])-T_i([w,\ba^{(t)}_{-i}]).\label{eqEstimatedUtility}
\end{align}}
Both terms in \eqref{eqEstimatedUtility} can be computed by BS $w$
and fed back to user $i$. To compute the first term in
\eqref{eqEstimatedUtility}, BS $w$ solves its resource allocation
problem with the set of users $\mathcal{N}_w([w,\ba^{(t)}_{-i}])$.
To obtain the second term in \eqref{eqEstimatedUtility}, BS $w$
solves its per-cell problem with and without user $i$ (cf.
\eqref{eqUtilitySimple}). Again only local channel states are
needed. It is important to note that to carry out the proposed
mechanism, the users do not need to know the behaviors of their
counterparts. They only need the ``summary" information (the tax and
the rate estimates) from the BSs.



In practice, the users may only switch to a new BS {\it if it offers
significantly higher utility}, because each of such switch induces
costs such as message passing. Let us use $c_i$ to denote such cost
for user $i$. When switching costs are included into the decision
process, in each iteration of the mechanism, $w^*\in
BR_i(\mathbf{a}^{(t)})$ implies $U_i([w^*,\mathbf{a}^{(t)}_{-i}])>
U_i(\mathbf{a}^{(t)})+c_i$. This modification could reduce the
number of iterations needed for convergence (since the users are now
less willing to change association), but could also reduce the
system throughput achieved by the identified NE.


The convergence property of the proposed mechanism is provided in
the following theorem. The details of the proof is relegated to the
Appendix.
\newtheorem{T7}{Theorem}
\begin{T1}\label{theoremConvergence}
{\it When choosing $M\ge N$, the BS association mechanism produces a
sequence $\left\{\mathbf{a}^{(t)}\right\}_{t=1}^{\infty}$ that
converges to a NE of game $\mathcal{G}$ with probability 1 (w.p.1).}
\end{T1}

%

\vspace{-0.2cm}
\section{Discussions  and Extensions}\label{secDiscussion}
To this point we have assumed that the BSs are interested in
maximizing the per-BS throughput. Such assumption allows the BSs to
have closed-form solution to their optimization problems, and it
leads to important properties such as submodularity of the
throughput functions. Our work can be extended to cases where the
BSs allocate resources using general utility functions as well.

First we mention that all the previous properties of the mechanism
can be straightforwardly generalized to the case where each BS $w$
aims to maximize a {\it weighted} throughput of the form
$\sum_{i\in\mathcal{N}_w}{\gamma_i}r_i$. The set of weights
$\{\gamma_i\ge 0\}_{i=1}^{N}$ can be adjusted adaptively by the BSs
over time to ensure fairness among the users' {\it time-averaged}
transmission rates (see e.g., \cite{yu11}).

Consider an alternative case in which BS $w$ is interested in
finding the best channel assignment to achieve the proportional
fairness (PF). The per-BS problem is then given by
\cite{song05b}{\small
\begin{align}
\max_{{\boldsymbol{\beta}_w }}&\quad
\sum_{i\in\mathcal{N}_w(\mathbf{a})}\alpha_w\log\left(r_i(\bfbeta,\mathbf{p},\mathbf{a})\right) \nonumber\tag{CA-PF}\\
{\textrm{s.t.}} &\quad
\beta^k_w\in\mathcal{N}_w(\mathbf{a}),\forall~k\in\mathcal{K}_w.\nonumber
\end{align}}
This problem generally does not admit a closed-form solution, and
the BS needs to perform numerical search to obtain the optimal
solutions (see \cite{song05b} for a set of efficient search
algorithms). Let us use $r^{\rm PF}_i(\mathbf{a})$ to denote the
resulting transmission rate for user $i$. Following
\eqref{eqUtilitySimple}, each user $i$ in cell $w$ has the following
utility {\small $U^{\rm PF}_i(\mathbf{a})\triangleq
\alpha_w\log(r^{PF}_i(\mathbf{a}))-T^{PF}_i(\mathbf{a})$}, where
{\small $
 T^{\rm PF}_i(\mathbf{a})\triangleq \alpha_w\sum_{ j\in\mathcal{N}_{w}({\mathbf{a}}_{-i})}
\log(r^{\rm PF}_j({\mathbf{a}}_{-i}))-\alpha_w\sum_{
j\in\mathcal{N}_{w}(\mathbf{a})\setminus i}\log(r^{\rm
PF}_j(\mathbf{a}))$}. We can now construct a PF association game
$\mathcal{G}^{\rm PF}$ with each user's utility function given as
$U^{\rm PF}_i(\cdot)$. Similarly as in Theorem
\ref{theoremExistenceS}, we can show that the optimal association
profile {$ \mathbf{a}^*=
\arg\max_{\mathbf{a}}\sum_{w}\alpha_w{\sum_{i\in\mathcal{N}_w(\mathbf{a})}\log(r^{\rm
PF}_i(\mathbf{a}))} $} must be a NE of this game. Our proposed
mechanism can be applied for finding the
NE of this game. 

In general, most of the properties of the mechanism (except for the
efficiency lower bounds of the NE) can be extended to networks with
the following properties: 1) The BSs operate independently (using
orthogonal time/frequencey resources); 2) The utility function
chosen by each BS is {\it separable} among the associated users; 3)
The optimal solutions of the per-BS optimization problem can be
found. However, it is not clear at this point whether the
monotonicity and the submodularity conditions can be carried over to
the case of general utility functions. Showing such properties under
general utility functions is left as a future work.

\vspace{-0.0cm}
\section{Simulations}\label{secSimulation}
In this section, we present simulation results to demonstrate the
performance of the proposed algorithm. Both indoor and outdoor
network scenarios are considered.

\vspace{-0.3cm}
\subsection{An Indoor Network Scenario}
\begin{figure*}[htb] \vspace*{-0.4cm}
\begin{minipage}[4]{0.49\linewidth}
    \centering
    \vspace*{-0.2cm}
    {\includegraphics[width=
1\linewidth]{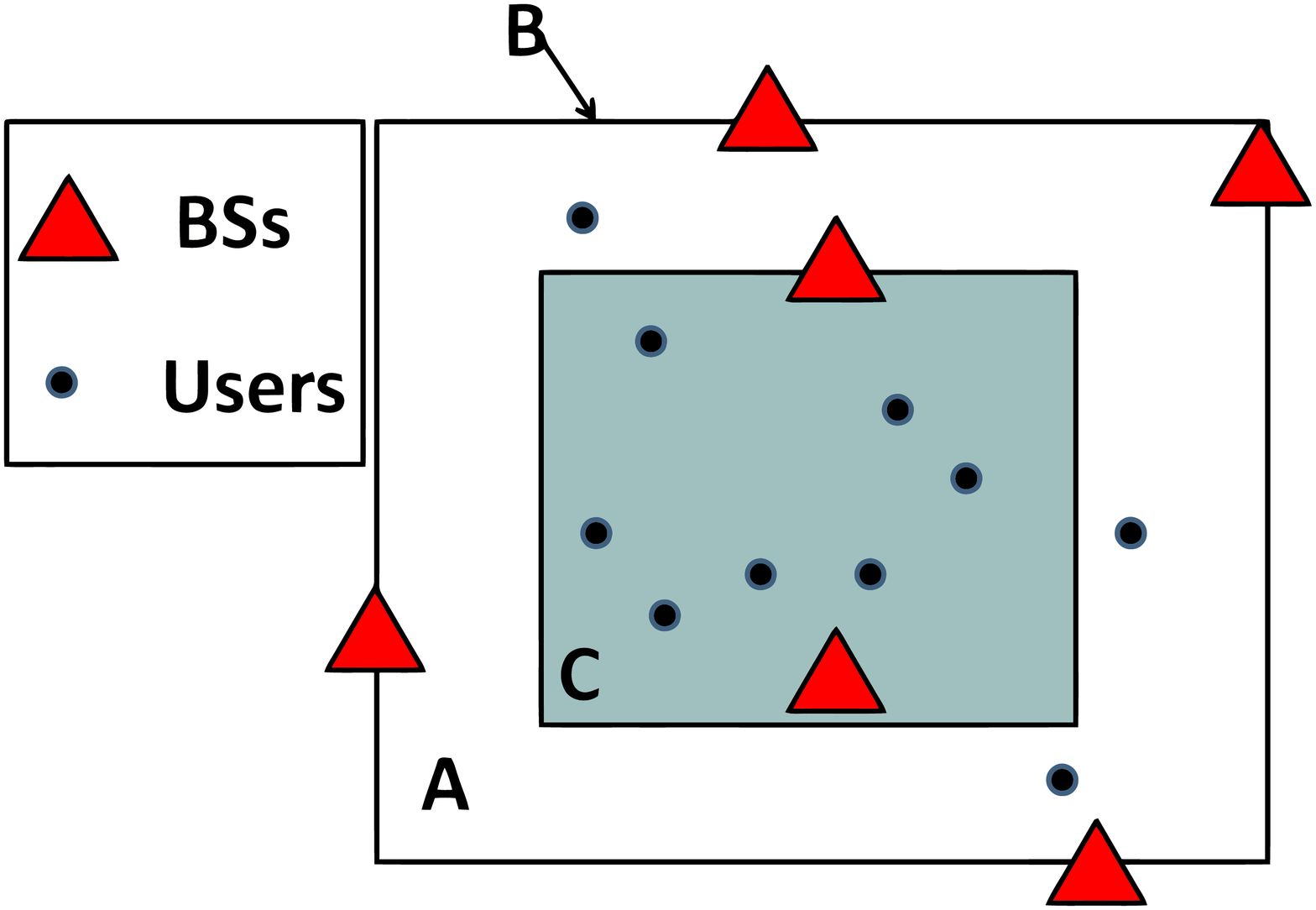}
\vspace*{-1cm}\caption{Illustration of the simulation setting with
$N=10$, $W=6$, $D=0.3$. }\label{figSchematic} \vspace*{-0.1cm}}
\end{minipage}\hfill
\begin{minipage}[4]{0.49\linewidth}
    \centering
    {\includegraphics[width=
1\linewidth]{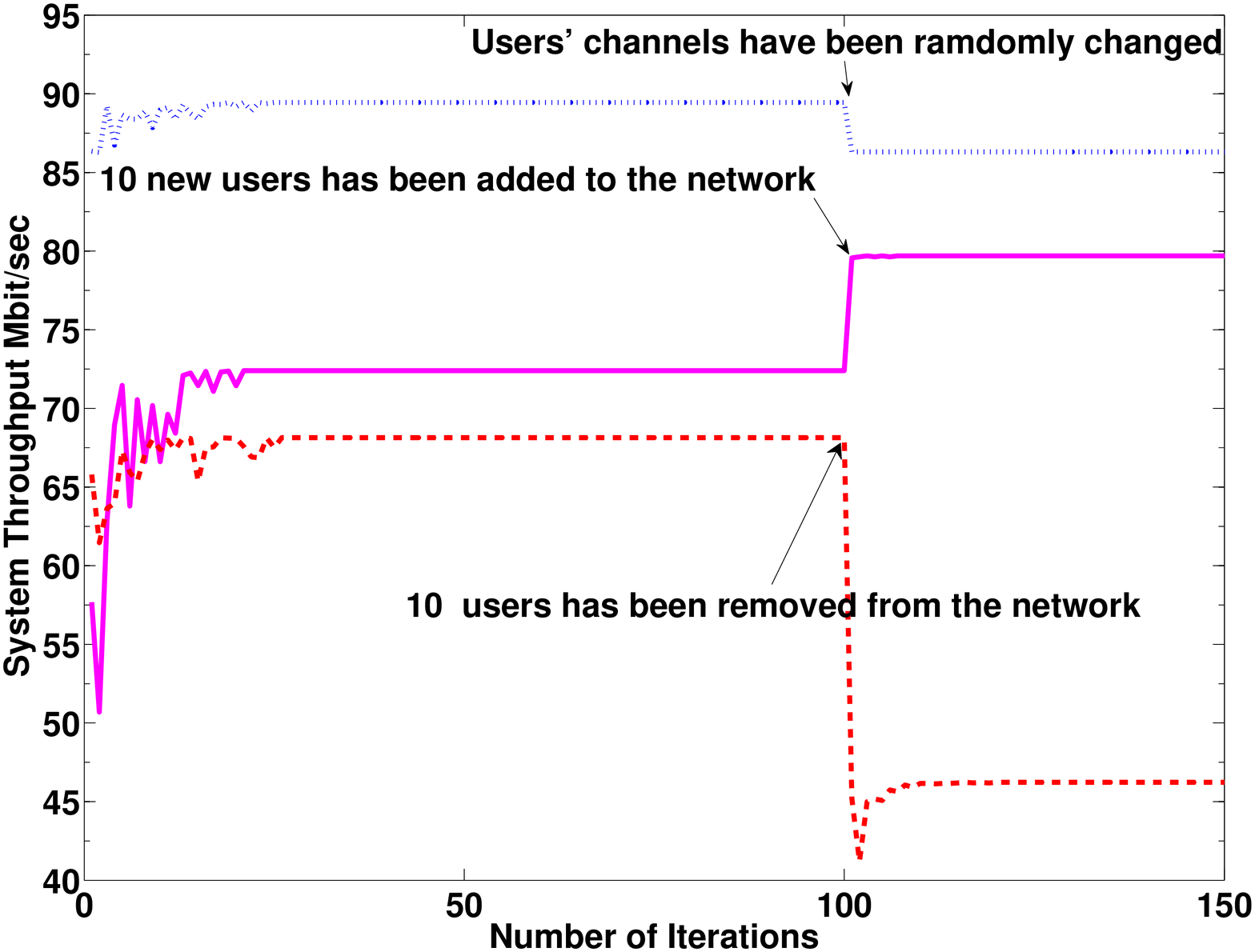}
\vspace*{-0.6cm}\caption{Three realizations of throughput. $K=512$,
$N=20$, $W=8$. 
}\label{figRealizationTrack} \vspace*{-0.1cm}}
\end{minipage}
\vspace*{-0.5cm}
    \end{figure*}

We have the following settings for this part of the simulation. Let
us denote a $50m\times50m$ indoor area as $A$; denote the $25m\times
25m$ central area of $A$ as $C$; define the border of $A$ as $B$.
Define the parameter $0\le D \le1$ as the {\it distribution factor}
of the users/BSs: 1) $D\times 100\%$ of the users and BSs are
randomly placed in $A$; 2) the rest of the users are randomly placed
in $C$ and the rest of the BSs are randomly placed on $B$. When $D$
is small, the subset of BSs that are located at the center of the
area become hotspots and are likely to be congested. See Fig.
\ref{figSchematic} for an illustration. Let $d_{i,w}$ denote the
distance between user $i$ and BS $w$. The channels between user $i$
and BS $w$,  $\{h^k_{i}\}_{k\in\mathcal{K}_w}$, are generated
independently from the complex Gaussian distribution
$\mathcal{CN}(0, \sigma^2_{i,w})$, with
$\sigma^2_{i,w}=L_{i,w}/PL_{i,w}$. The random variable $L_{i,w}$
models the shadowing effect, i.e., $10\log10(L_{i,w})\sim
\mathcal{N}(0,64)$ is a real Gaussian random variable. The variable
$PL_{i,w}$ is the pathloss between BS $w$ and user $i$. To model the
pathloss in the indoor scenario, the office environment model
\cite{rappaport02} is used. The key simulation
parameters are given in Table \ref{tableSimulation}. 
The performance of the proposed algorithm will be compared with the
algorithm that first assigns the users to their nearest BSs, and
then optimally perform the per-BS resource allocation. Note that
this algorithm {\it separates} the process of association and
per-cell resource allocation, hence in most cases gives degraded
system performance. Throughout this subsection, the CAPA strategy
will be adopted for per-BS resource allocation.

\vspace*{-0.1cm}
\begin{table*}[htb]
\caption{\small Simulation Parameters for the Indoor
Network.}\vspace*{-0.3cm}
\begin{center}
{\small
\begin{tabular}{|c |c | c|c|}
\hline
{\bf Parameters} & {\bf Values} & {\bf Parameters} & {\bf Values}\\
 \hline
{\bf $\bar{p}_w$}& $23$ dBm &{\bf Total Bandwidth}& $80$MHz \\
 \hline
{\bf Pass Loss (dB)} &$ {\rm PL}_{i,w}={\rm PL}(1)+26\log10(\frac{d_{i,w}}{1})+14.1 $& {\bf $\alpha_w$} & $1$\\
 \hline
{\bf BER} & $10^{-6}$&  {\bf Length of Memory} & $10$\\
 \hline
{\bf Operating Frequency} & $1.9$ GHz &{\bf Noise Power} & $-100$ dBm/Hz\\
 \hline
\end{tabular} } \label{tableSimulation}
\end{center}
\vspace*{-0.1cm}
\end{table*}

The first set of experiments evaluate the convergence performance of
the proposed
mechanism. 
Fig. \ref{figRealizationTrack} plots 3 realizations of the evolution
of system throughput. This figure demonstrates the ability of the
algorithm to ``track" the equilibrium solutions. The algorithm takes
a few iterations to converge to new equilibria when the following
events occur at iteration $100$: 1) $10$ (randomly placed) new/old
users enter/leave the system; 2) all of the users' channel gains are
re-generated (with the locations of the users and BSs unchanged).

\begin{figure*}[htb]
\vspace*{-.3cm}
    \begin{minipage}[t]{0.49\linewidth}
    \centering
    {\includegraphics[width=
1\linewidth]{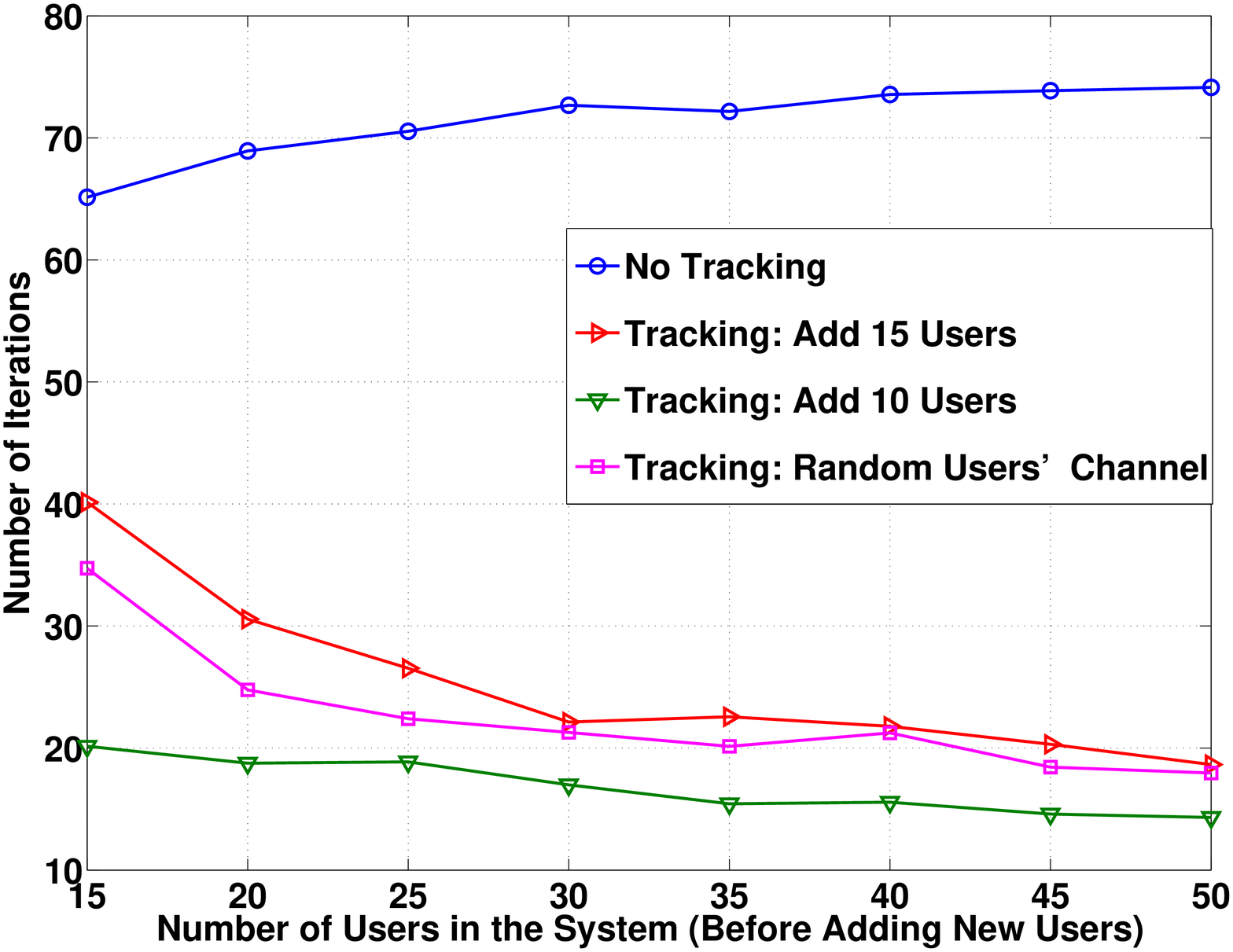}
\vspace*{-0.8cm}\caption{Averaged convergence time v.s. number of
users. Each point in this figure is averaged over $200$ random
networks. $W=8$, $K=512$,
$D=0.4$.}\label{figSpeedTrack}\vspace*{-0.1cm}}
\end{minipage}\hfill
    \begin{minipage}[t]{0.49\linewidth}
    \centering
    {\includegraphics[width=
1\linewidth]{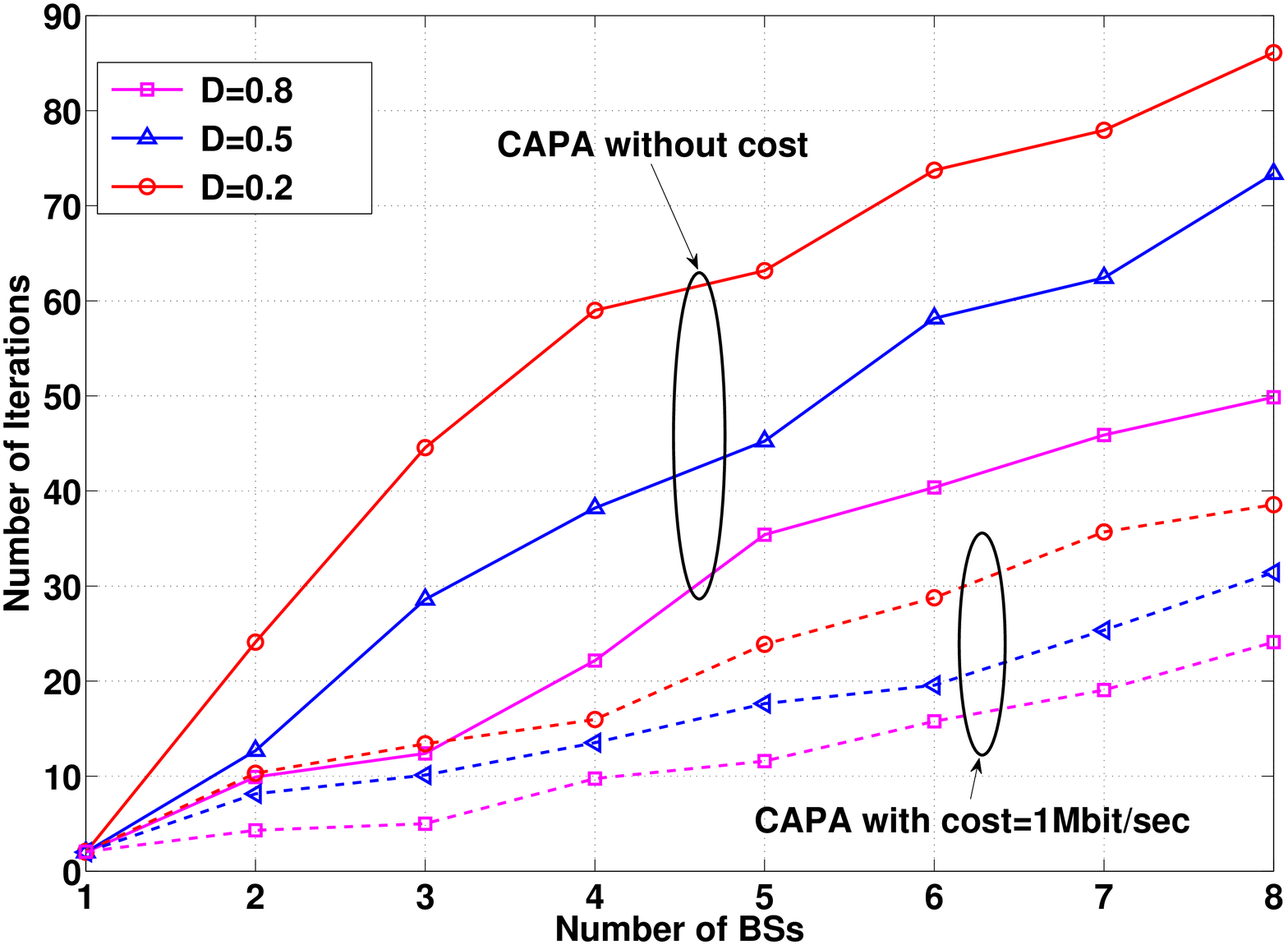} \vspace*{-0.8cm}\caption{
Averaged convergence time v.s. number of BSs with and w/o costs.
Each point in this figure is averaged over $200$ random networks.
$c_i=1$ Mbit/sec, $N=30$, $K=512$,
$D=\{0.2,~0.5,~0.8\}$.}\label{figSpeedD} \vspace*{-0.1cm}}
\end{minipage}
\vspace*{-0.4cm}
    \end{figure*}

In Fig. \ref{figSpeedTrack}, we evaluate the averaged convergence
time for the algorithm. We highlight its ``tracking" ability by
adding a number of new users and by randomly re-generating all the
users' channel gains {\it after an equilibrium has been reached}.
The algorithm is able to track the equilibrium much faster than
performing a complete restart. In Fig. \ref{figSpeedD}, we plot the
averaged convergence time of the algorithm with $N=30$, $K=512$. We
observe that the convergence time is decreasing with $D$. This
phenomenon is intuitive because when $D$ is close to $1$, the event
of congestion is less likely to happen as on average the
communication load is evenly distributed among the BSs. On the
contrary, when $D$ becomes small, those BSs located in the interior
of the area are likely to be congested. Large portions of the users
will then seek for alternative choices, which results in longer
convergence time. Additionally, when taking the switching costs into
consideration, the algorithm converges significantly faster.

%
%
%

\begin{figure*}[htb] \vspace*{-.5cm}

\begin{minipage}[t]{0.49\linewidth}
\centering
    {\includegraphics[width=
1\linewidth]{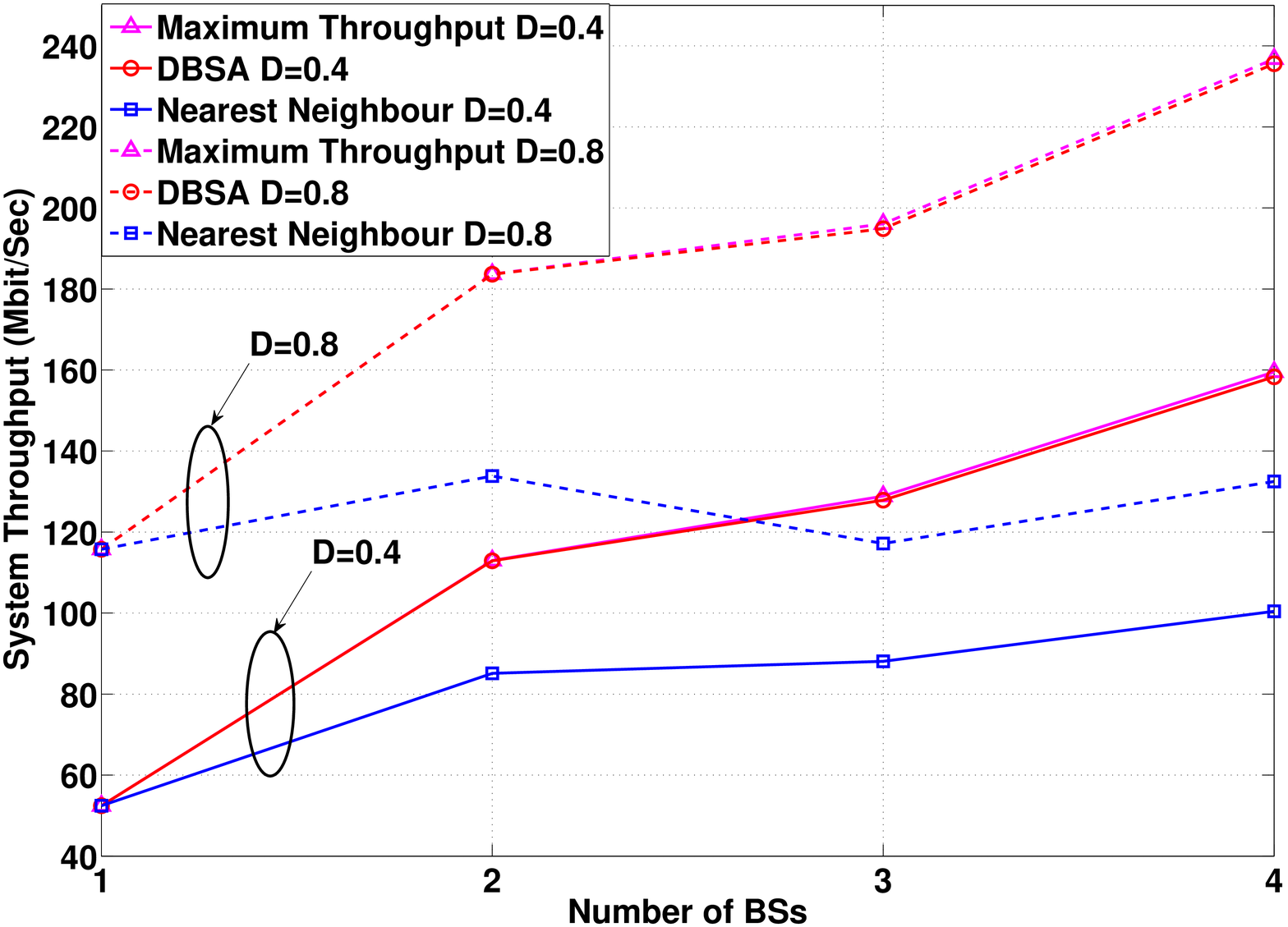}
\vspace*{-0.8cm}\caption{Averaged system throughput v.s. number of
BSs by different algorithms and the maximum achievable throughput.
Each point in this figure is averaged over $100$ random networks.
$N=10$, $K=64$, $D=\{0.4,~0.8\}$.}\label{figExhaustive}
\vspace*{-0.1cm}}\end{minipage}\hfill
    \begin{minipage}[t]{0.49\linewidth}
    \centering
    {\includegraphics[width=
1\linewidth]{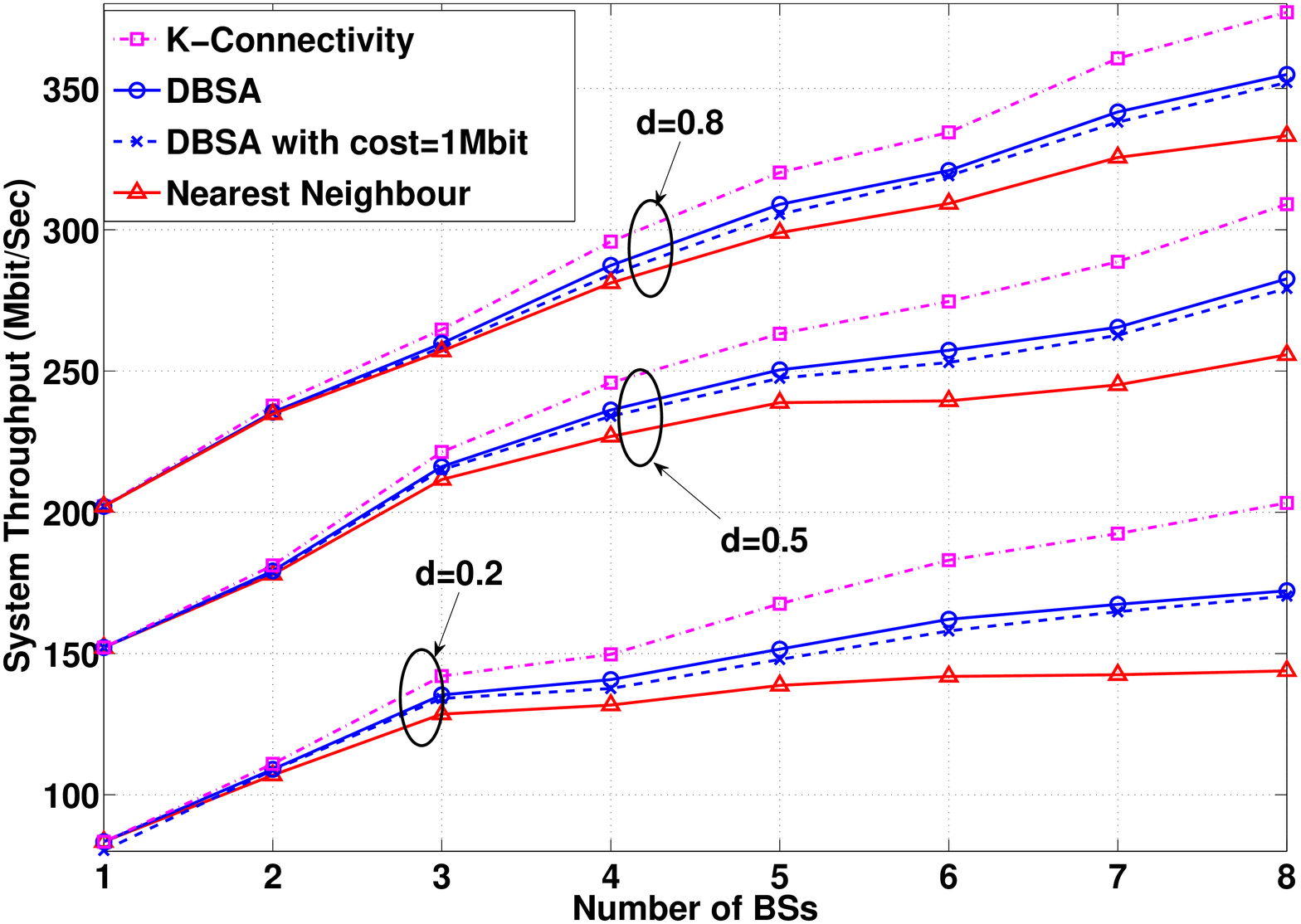} \vspace*{-0.8cm}\caption{Averaged
system throughput v.s. number of BSs by different algorithms.
$N=30$, $K=512$, $D=\{0.2,~0.5,~0.8\}$. Each point in this figure is
averaged over $200$ random networks.
}\label{figRateD}\vspace*{-0.1cm}}
\end{minipage}

\vspace*{-0.1cm}
    \end{figure*}

The second set of experiments intend to evaluate the throughput
performance of the proposed algorithm. We first investigate a
relatively small network with $10$ users, $64$ channels and $1-4$
BSs, and compare the performance of the proposed algorithms
to the global optimal solution of the problem (SYS) (obtained by an exhaustive search). 
The results are shown in Fig. \ref{figExhaustive}. We see that the
proposed algorithm, abbreviated as Distributed BS Association
(DBSA), performs well with little throughput loss. In contrast, the
nearest BS algorithm performs poorly.

We then evaluate the performance of the algorithm in larger networks
with $30$ users, up to $8$ BSs and $512$ channels. Fig.
\ref{figRateD} shows the comparison of the averaged performance of
the proposed algorithm and the nearest BS algorithm. Due to the
prohibitive computation time required, we are unable to obtain the
optimal system throughput in this case. We instead compute a
(strict) upper bound of the maximum throughput assuming that the
users can connect to multiple BSs {\it simultaneously}. We refer to
this as the {\it multiple-connectivity} network. We also observe
that when we take the switching costs into consideration ($c_i=1$
Mbps for all $i$), there is a slight decrease in system throughput.

\begin{figure*}
    \begin{minipage}[t]{0.49\linewidth}
\centering
    {\includegraphics[width=
1\linewidth]{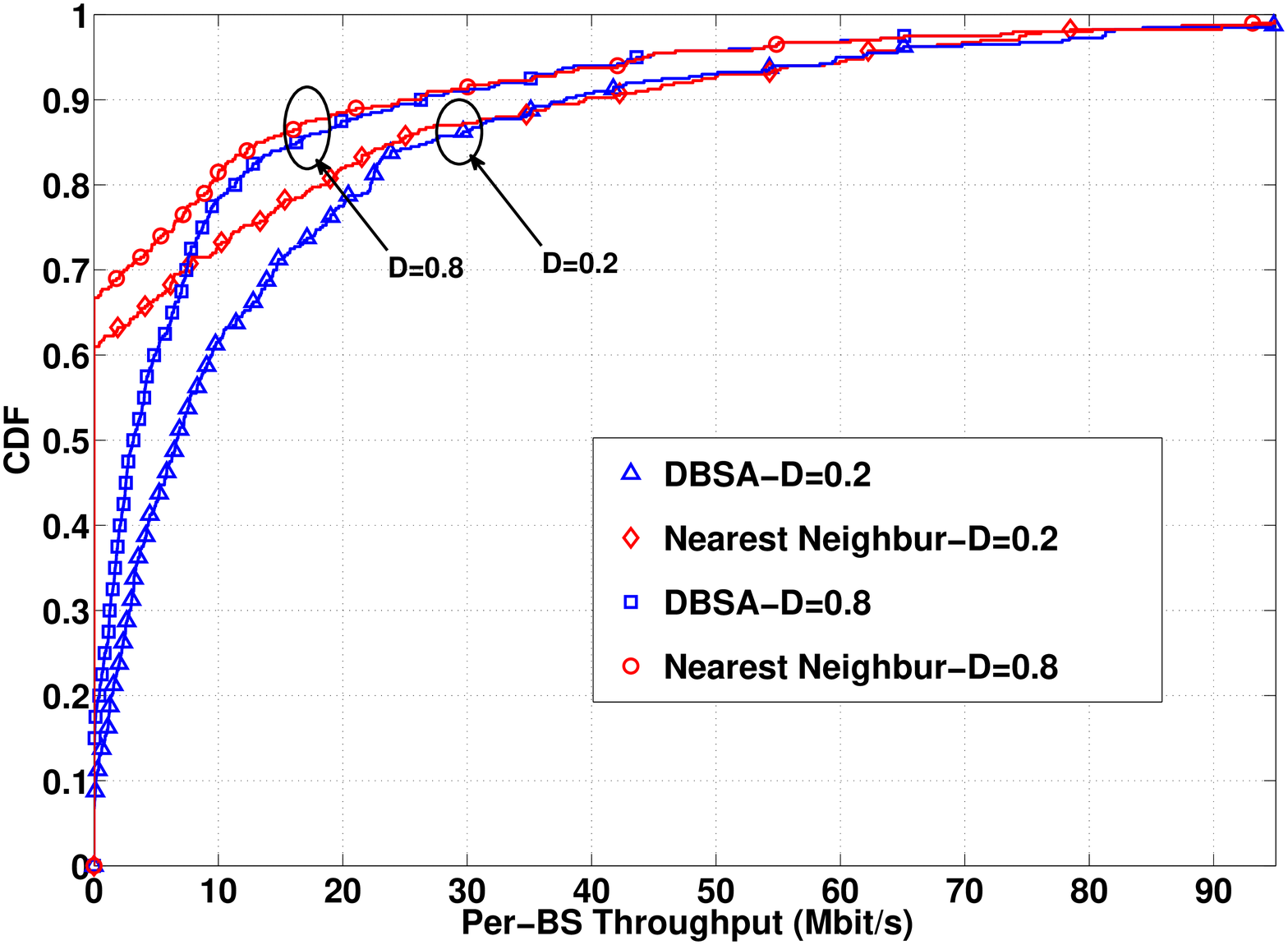}
\vspace*{-0.8cm}\caption{Empirical CDF of the per-BS rate. Each
curve in this figure composed of the rate of the BSs over $100$
random networks. $W=8$, $N=30$, $K=512$,
$D=\{0.2,~0.8\}$.}\label{figBSRateDistribution}
\vspace*{-0.1cm}}\end{minipage}
\end{figure*}

In Fig. \ref{figBSRateDistribution}, we show the distribution of the
per-BS rates achieved by the proposed algorithm and the nearest BS
algorithm. From the figure we see that the proposed algorithm is
able to distribute the throughput to different BSs fairly, while the
nearest BSs algorithm may result in severe unbalance of the BSs'
loads (some BSs may experience heavy traffic while the rest of the
BSs may become idle).

\vspace{-0.2cm}
\subsection{An Outdoor Multicell Cellular Network Scenario}
In this section we demonstrate the performance of the proposed
algorithm in a multicell OFDMA cellular network. Standard cellular
network parameters are used for the simulation, see Table
\ref{tableSimulationOutDoor} \footnote{Most of the network
parameters are taken from \cite{yu11}. In the present work only
single antenna systems are simulated. }. Again frequency selective
channels with a Rayleigh fading component and $8$ dB log-normal
fading component are simulated. Users are assumed to be distributed
uniformly in the entire network. Throughout this subsection, the
system level PF objective is optimized, thus the CA-PF strategy
discussed in Section \ref{secDiscussion} is used for the per-BS
resource allocation. The solution to the problem (CA-PF) is computed
using Algorithm 1 in \cite{song05b}. The main purpose of this
experiment is to evaluate the performance of the proposed algorithm
when some of the key assumptions guaranteeing the theoretical
properties of the algorithm no longer hold true. Note that in order
for the proposed algorithm to work in this network setting,
inter-cell interference should be treated as noise. That is, user
$i$'s  noise power on channel $k$, $n^k_i$, should include both the
environmental noise power {\it and} the inter-cell interference
power.

\vspace{-0.1cm}
\begin{table}[h]
\caption{Simulation Parameters for the Outdoor
Network}\vspace*{-0.3cm}
\begin{center}
{\small
\begin{tabular}{|c |c |}
\hline
{\bf Parameters} & {\bf Values}\\
\hline Cell layout & Hexagonal, 7 cells, 3 sectors per cell\\
\hline
 BS-BS distance& 2.8 km\\
 \hline
 Frequency Reuse & 1\\
 \hline
$\bar{p}_w$& $49$ dBm\\
 \hline
Pass Loss Model (dB) &$ {\rm PL}_{i,w}=128.1+36.7\log10(d_{i,w}) $\\
 \hline
BER & $10^{-6}$\\
 \hline
Total Bandwidth& $10$ MHz\\
 \hline
Noise Power & $-169$ dBm/Hz\\
\hline
    Multipath Time Delay Profile & ITU-R M.1225 PedA\\
 \hline
 Number of channel (FFT size)& 64\\
 \hline
Length of Individual Memory & $M=10$\\
 \hline
\end{tabular} } \label{tableSimulationOutDoor}
\end{center}
\vspace{-0.2cm}
\end{table}
We first show the convergence of the algorithm. In the considered
cellular network, different BSs transmit using {\it the same}
spectrum bands. Consequently our theoretical analysis of the
convergence is no longer valid. However, convergence is still
observed empirically. See Table \ref{tableSpeedMulticell} for the
comparison of the convergence speed with and without the switching
costs $\{c_i\}$.

\begin{figure*}[htb] \vspace*{-.2cm}
    \begin{minipage}[t]{0.49\linewidth}
    \centering
    {\includegraphics[width=
1\linewidth]{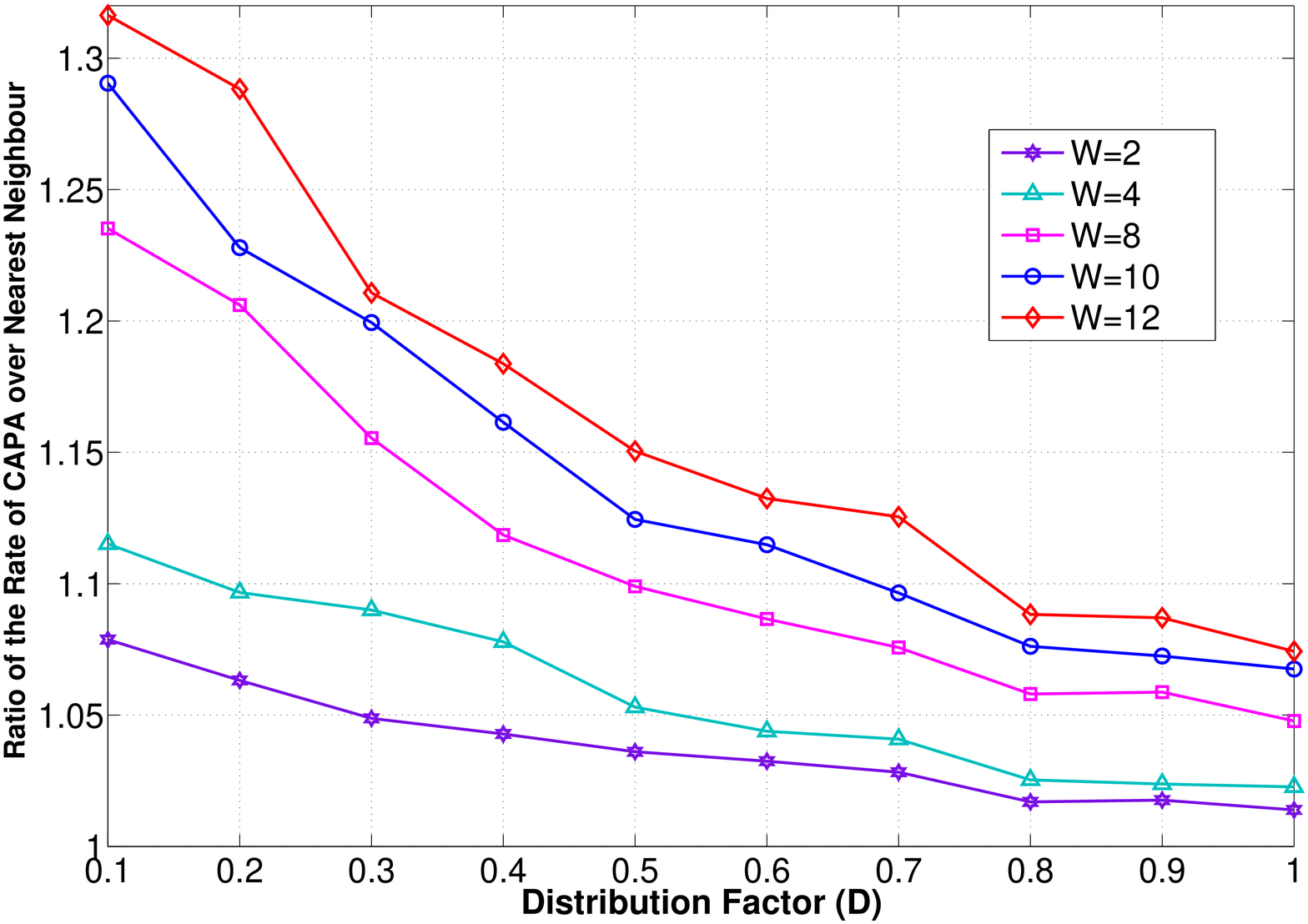} \vspace*{-0.8cm}\caption{Averaged
throughput ratio (CAPA over nearest neighbor) v.s. distribution
factor D. $N=30$, $K=512$, $W=\{2, ~4,~8,~10,~12\}$. Each point in
this figure is averaged over $200$ random networks.
}\label{figRatio} \vspace*{-0.1cm}}
\end{minipage}\hfill
    \begin{minipage}[t]{0.49\linewidth}
    \centering

    {\includegraphics[width=
1  \linewidth]{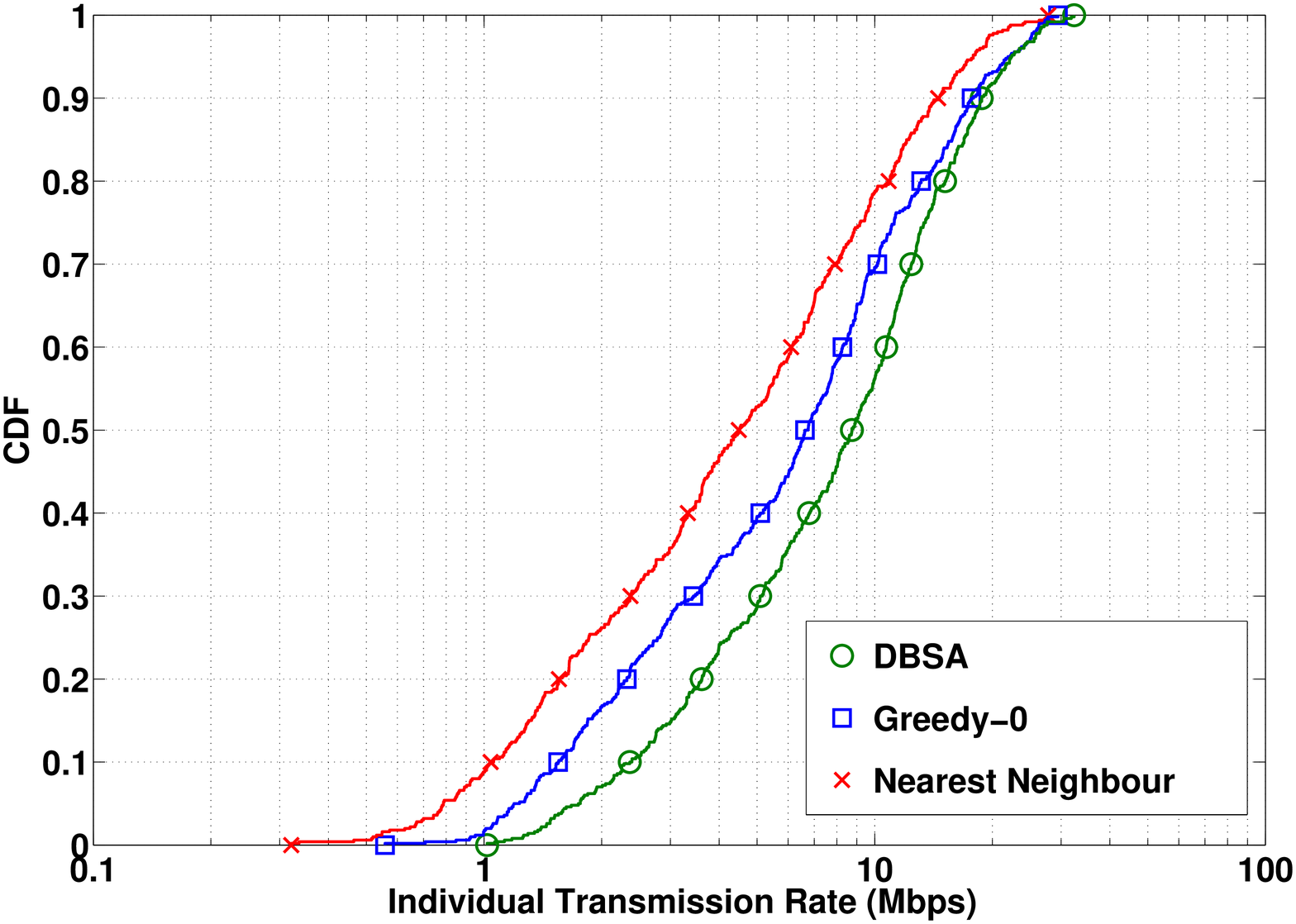}
\vspace*{-0.8cm}\caption{Comparison of the empirical CDF of the
users' rates in the multicell cellular networks. $N=40$. Each curve
in this figure is the CDF of the users' rates in $100$ generations
of the network.}\label{cdfMulticell} \vspace*{-0.1cm}}

\end{minipage}
\vspace*{-0.3cm}
    \end{figure*}

We then demonstrate the throughput performance of the algorithm. We
compare the proposed algorithm with the nearest BS algorithm and the
``Greedy-0" algorithm proposed in \cite{bu06}, which is a
centralized algorithm that finds a good user-BS association by
successively perturbing the user-BS association locally. In Table
\ref{tableRateMulticell} and Fig. \ref{cdfMulticell}, we see that
the proposed algorithm compares favorably with the other algorithms
both in terms of system throughput and fairness levels. Each entry
in the table is obtained via an average of $200$ randomly generated
networks.
\vspace*{-0.1cm}
\begin{table*}[htb]
\caption{ The Averaged Number of Iterations for
Convergence}\vspace*{-0.5cm}
\begin{center}
{\small
\begin{tabular}{|c |c |c|c|}
\hline
& DBSA & DBSA $c_i=0.1$ Mbps& DBSA $c_i=0.5$ Mbps\\
 \hline
 N=10& 55 & 33 & 21 \\
 \hline
 N=30& 65 & 38 & 25 \\
 \hline
 N=50 & 70 & 40 & 23 \\
 \hline
\end{tabular} } \label{tableSpeedMulticell}
\end{center}
\vspace*{-0.5cm}
\end{table*}

\begin{table*}[htb]
\caption{ Comparison of the System Throughput of Different
Algorithms}\vspace*{-0.3cm}
\begin{center}
{\small
\begin{tabular}{|c |c|c|c |c|c|}
\hline
& DBSA & DBSA ($c_i=0.1$ Mbps)& DBSA ($c_i=0.5$ Mbps) &Greedy-0& Nearest\\
 \hline
 N=20& 97.86 Mbps& 93.08 Mbps& 90.13 Mbps& 82.23 Mbps& 63.31 Mbps\\
 \hline
 N=40& 117.9 Mbps& 115.1 Mbps& 109.3 Mbps& 105.7 Mbps& 89.0 Mbps\\
 \hline
 N=60 & 135.9 Mbps& 129.9 Mbps& 125.1 Mbps& 119.5 Mbps& 104.8 Mbps\\
 \hline
\end{tabular} } \label{tableRateMulticell}
\end{center}
\vspace{-0.2cm}
\end{table*}

\begin{figure*}[htb] \vspace*{-.2cm}
    \begin{minipage}[t]{0.49\linewidth}
    \centering
  {\includegraphics[width=
1\linewidth]{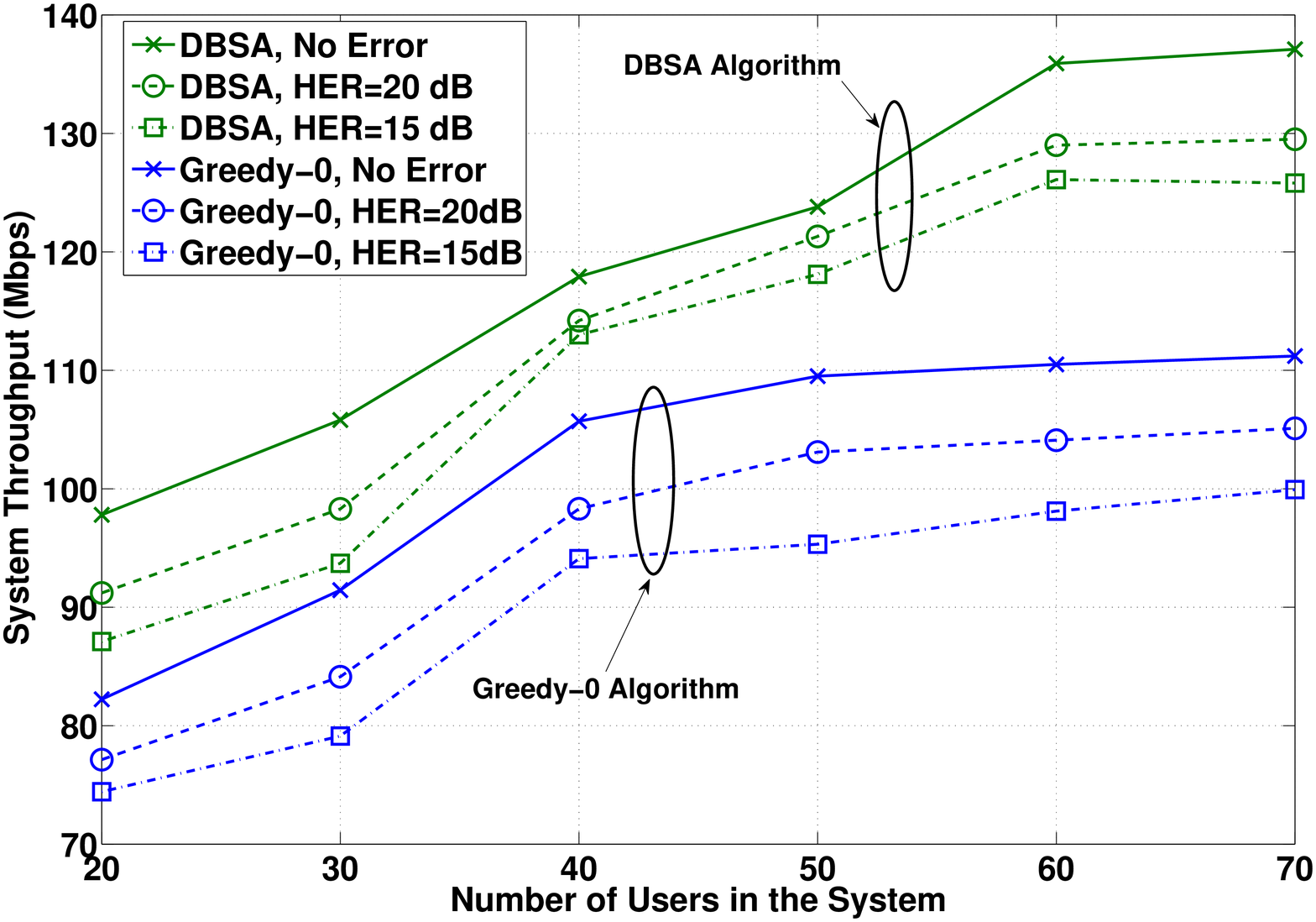}
\vspace*{-0.8cm}\caption{Evaluation of the performance of the
proposed algorithm in the presence of channel estimation error. Each
point in this figure is averaged over $200$ random
networks.}\label{figRateInaccurate} \vspace*{-0.1cm}}
\end{minipage}\hfill
    \begin{minipage}[t]{0.49\linewidth}
    \centering
    {\includegraphics[width=
1  \linewidth]{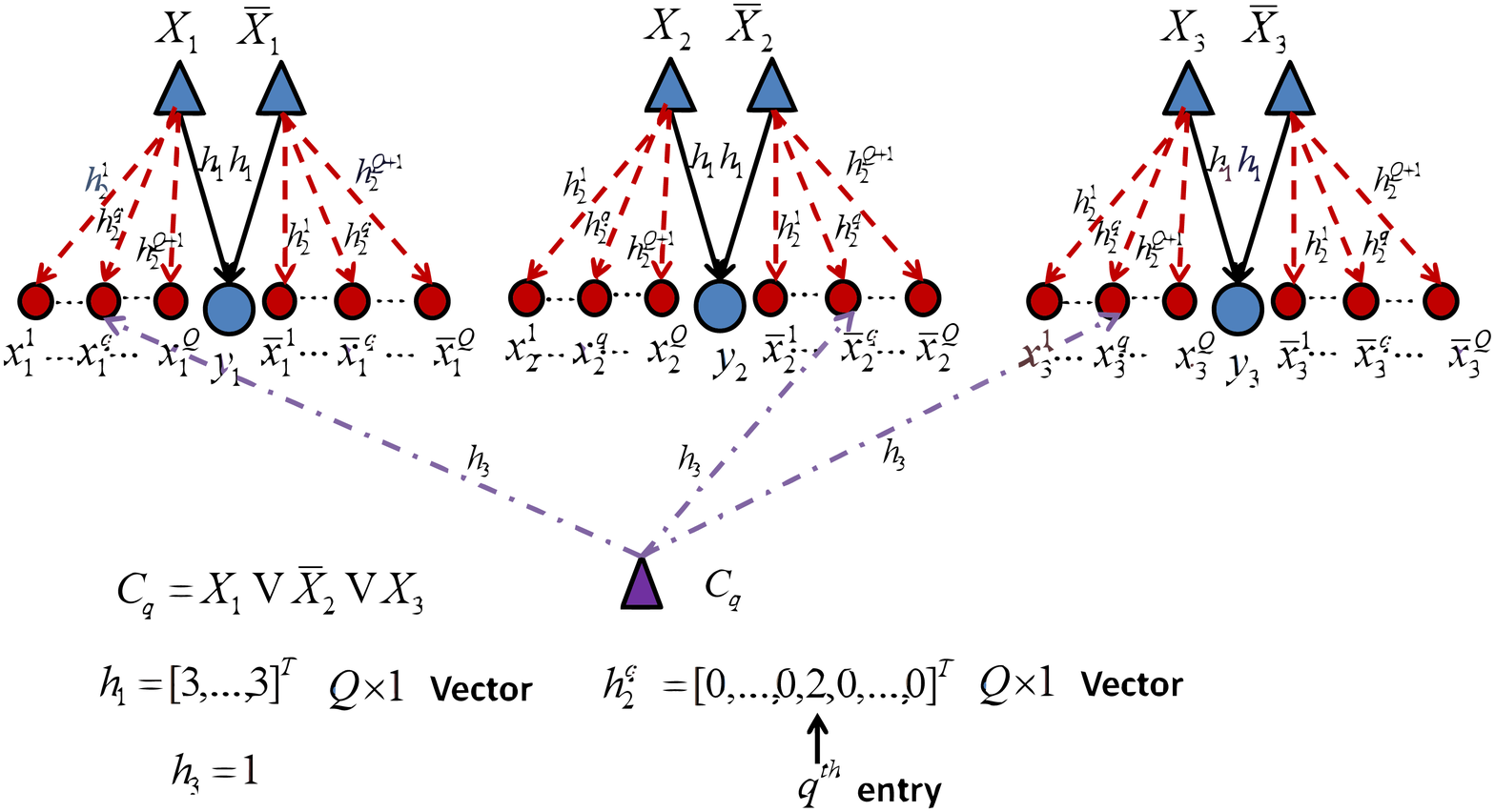}
\vspace*{-0.8cm}\caption{Construction of the network for clause
$C_q=X_1 \vee\bar{X}_2 \vee X_3$ for CAPA resource allocation
strategy. }\label{figConstructionCAPA} \vspace*{-0.1cm}}
\end{minipage}
\vspace*{-0.3cm}
    \end{figure*}

Additionally, we evaluate the performance of the algorithm when only
noisy channel information is available. In particular, we consider
the situation in which the {\it normalized channel magnitude}
estimated by the users is subject to a zero mean estimation error
\cite{benedict67,pauluzzi00}. Let
$|\widetilde{h^k_{i}}|^2\triangleq\frac{|{h^k_{i}}|^2}{n^k_i}+\epsilon^k_i$
denote the estimated normalized channel magnitude by user $i$ on
channel $k$, where $\epsilon^k_i\sim\mathcal{N}(0,(\sigma^k_i)^2)$
is the estimation error. Suppose the estimated normalized channels
are used by the BSs for resource allocation. To evaluate the
performance loss due to the channel inaccuracy, similarly as in
\cite{hong11a}, we introduce a term called Channel Error Ratio (CER)
to quantify the strength of the channel error: $ {\rm
CER}^k_i\triangleq
10\log_{10}\left(\frac{|h^k_i|^2}{n^k_i(\sigma^k_i)^2}\right)$. In
Fig. \ref{figRateInaccurate}, we plot the system throughput
performance with different values of the CER. We observe that the
overall throughput degrades slightly with such inaccurate channel
information.



\vspace{-0.2cm}
\section{Conclusion}\label{secConclusion}
In this work, we studied a resource management problem in a
multi-cell network in the presence of strategic/selfish users. We
propose a novel mechanism that implements a strategy-proof and
approximately optimal scheme in dominant strategies. Utilizing a key
submodularity property of the per-BS throughput function, we
characterized the efficiency of the proposed
mechanism. 
As a future work, we will study the case in which there is limited
(low rate) feedback from the users to the BSs. In this case feedback
strategy needs to be designed in conjunction with the BSs' and the
users' strategies. A new approximation ratio needs to be derived for
this more practical scenario. We also plan to generalize the
submodularity property to other utility functions and to the case of
MIMO networks. Another interesting extension of the current work is
to include the users that are {\it hostile} instead of {\it
non-cooperative}. Strategies that are different from pricing are
needed in this case to counter the untruthfulness induced by the
hostility.

\vspace{-0.3cm}
\section{Acknowledgement}
The authors wish to thank Prof. Tom Luo for helpful discussion and
various suggestions.

\setstretch{1} \vspace{-0.2cm}

\section{Appendix}
\vspace{-0.2cm}
\subsection{Proof of Theorem \ref{theoremNP}}\label{appNP} Without
loss of generality, we consider the case where $\alpha_w=1,\
n^k_i=1, \forall~w,k,i$. This claim is proved based on a polynomial
time transformation from 3-SAT problem, which is a known NP-complete
problem. The 3-SAT problem is described as follows. Given $Q$
disjunctive clauses $C_1\cdots,C_Q$ defined on $M$ Boolean variables
$X_1\cdots,X_M$, i.e., $C_q=T_q^1\vee T_q^2\vee T_q^3$ with
$T_q^j\in\{X_1,\cdots,X_M, \bar{X}_1,\cdots,\bar{X}_M\}$, the
question is to check whether there exists a truth assignment for the
Boolean variables such that all clauses are satisfied. Define
$\pi(C_q)$ as the set of terms contained in clause $C_q$, and define
$\mathcal{I}(T)$ as the index of a term $T$'s corresponding
variable, i.e., if $C_m=\bar{X}_1\vee \bar{X}_2\vee X_4$, then
$\pi(C_m)=\{\bar{X}_1, \bar{X}_2, X_4\}$, and
$\mathcal{I}(\bar{X}_1)=1$.

Given any instance of 3-SAT problem with $Q$ clauses and $M$
variables, we construct an instance of the multi-BS network with
$(2Q+1)M$ users and $2M+Q$ BSs. For each variable $X_m$, we do the
following: 1) associate with it two {\it variable BSs}
$X_m,\bar{X}_m$; 2) associate with it $(2Q+1)$ {\it variable users}
$\{x_{m}^{1},\cdots,x_m^Q\}$,
$\{\bar{x}_{m}^{1},\cdots,\bar{x}_m^Q\}$ and $y_m$; 3) set the
individual maximum  power of the BSs $X_m,\bar{X}_m$ to be $Q$,
respectively; 4) let each of the BSs $X_m,\bar{X}_m$ operate on $Q$
channels. For each clause $C_q$ we do the following: 1) associate
with it a single BS $C_q$; 2) set the maximum  power of BS $C_q$ to
be $1$; 3) let BS $C_q$ operate on a single channel. Denote the
$k^{th}$ channel in the set $\mathcal{K}_w$ as $\mathcal{K}_w(k)$.
The channel gains between various users and BSs are given as
follows. {\small
\begin{align}
|h_{y_m}^k|^2&= \left\{ \begin{array}{ll}
3,&w=X_m~\textrm{or}~w=\bar{X}_m,~\forall~k\in\mathcal{K}_w\\
0,&\textrm{otherwise},
\end{array}\right.|h_{x^q_m}^k|^2&= \left\{ \begin{array}{ll}
2,&w=X_m,~k=\mathcal{K}_w(q)\\
1,&w=C_q,~X_m\in\pi(C_q),~\forall~k\in\mathcal{K}_w\\
0,&\textrm{otherwise},
\end{array}\right.\nonumber\\
|h_{\bar{x}^q_m}^k|^2&= \left\{ \begin{array}{ll}
2,&w=\bar{X}_m,~k=\mathcal{K}_w(q)\\
1,&w=C_q,~\textrm{with}~\bar{X}_m\in\pi(C_q),~\forall~k\in\mathcal{K}_w\\
0,&\textrm{otherwise}.
\end{array}\right.
\end{align}}
For example, for a clause ${\small C_q=X_1 \vee \bar{X}_2 \vee
X_3}$,
 the constructed network is shown in Fig.
 \ref{figConstructionCAPA}.

In the following, for notational simplicity, for a term
$T\in\pi(C_q)$, we will use the short hand notation $T$ to denote
its corresponding variable BSs (instead of $X_{\mathcal{I}(T)}$ or
$\bar{X}_{\mathcal{I}(T)}$), use the notation $y_{\mathcal{I}(T)}$
and $\{t^1_{\mathcal{I}(T)},\cdots, t^Q_{\mathcal{I}(T)}\}$ for its
corresponding variable users. Take an arbitrary clause $C_q$ and a
term $T\in\pi(C_q)$. Two important observations can be made at this
point. Both of them are the direct consequence of our network
construction, and we omit their proofs due to space limitation.

\newtheorem{O1}{Observation}
\begin{O1}\label{observation2}
If at the optimal solution, BS $T$ selects its corresponding
variable user $y_{\mathcal{I}(T)}$ for transmission, then the subset
of variable users
$\{t_{\mathcal{I}(T)}^{1},\cdots,t_{\mathcal{I}(T)}^Q\}$ will not be
selected by BS $T$ at the optimal solution. 
A direct consequence of this observation is that
the maximum throughput that the BS pair $T, \bar{T}$ can achieve is
$2Q+ Q\log 3$.
\end{O1}

\newtheorem{O3}{Observation}
\begin{O1}\label{observation3}
Assuming $T\in\pi(C_q)$, then the maximum throughput achievable by
the BSs $\{T,\bar{T},C_q\}$ is $2Q+Q\log 3+1$. This throughput is
achieved if and only if BS $T$ selects user $y_{\mathcal{I}(T)}$, BS
$\bar{T}$ selects users
$\{\bar{t}_{\mathcal{I}(T)}^{1},\cdots,\bar{t}_{\mathcal{I}(T)}^Q\}$
and BS $C_q$ selects the user $t_{\mathcal{I}(T)}^q$. 
\end{O1}

Our main claim is that the given 3-SAT instance is satisfiable if
and only if the constructed network achieves a throughput of at
least $Q+2MQ+MQ\log3$.

Suppose the 3-SAT problem is satisfiable. Then for each clause $C_q$
there is a term $T^*_q\in\pi(C_q)$ such that $T^{*}_q=1$.  Let the
variable BS $T^*_q$ select user $y_{\mathcal{I}(T^*_q)}$, let the
variable BS $\bar{T}_q^*$ select all users
$\{\bar{t}^1_{\mathcal{I}(T^*_q)},\cdots,\bar{t}^Q_{\mathcal{I}(T^*_q)}\}$,
and let the clause BS $q$ select the variable user
$t^q_{\mathcal{I}(T^*_q)}$. By Observation \ref{observation3}, the
total throughput of variable BSs $T^*_q$, $\bar{T}^*_q$ and the
clause Bs $C_q$ is $2Q+Q\log3+1$. As a result, the overall system
throughput is $2MQ+QM\log3+Q$.

Conversely, suppose a throughput of $2MQ+MQ\log 3+Q$ is achieved.
From Observation \ref{observation2} we see that the maximum
throughput for a variable BS pair $X_m,\bar{X}_m$ is $2Q+Q\log 3$.
As we have $M$ variable BS pairs and $Q$ clause BSs, then {\it each}
clause BS must achieve throughput $1$ in order for the system to
achieve the throughput $M(2Q+Q\log 3)+Q$. Observation
\ref{observation3} implies that this is only possible if for each
clause BS $C_q$, there is at least one clause BS, say
$T^*_q\in\pi(C_q)$ that transmits to the corresponding variable user
$y_{\mathcal{I}(T^*_q)}$. Set $\{T^*_q\}_{q=1}^{Q}$ all to $1$, then
each clause $C_q$ will be satisfied. Consequently, the resulting
3-SAT problem is satisfied.

\vspace{-0.2cm}
\subsection{Proof of Theorem
\ref{theoremExistenceS}}\label{appExistence}
We prove this theorem by contradiction. Suppose
${\mathbf{a}}^*\in\arg\max_{\mathbf{a}}R(\mathbf{a})$, but
${\mathbf{a}}^*$ is not a pure NE. Then there must exist a user $i$
such that $BR_i({\mathbf{a}}^*)\ne \emptyset$. Choose
$\widetilde{w}\in BR_i({\mathbf{a}}^*)$, and define a new
association profile
$\widetilde{\mathbf{a}}=[\widetilde{w},~{\mathbf{a}}^*_{-i}]$. Let
${w}^*={\mathbf{a}^*}[i]$. We show that user $i$'s unilateral change
of association has the same effect on its own utility as well as on
the system throughput {\small
\begin{align}
U_i(\widetilde{\mathbf{a}})-U_i({\mathbf{a}^*})& \stackrel{(a)}
=\sum_{j\in\mathcal{N}_{\widetilde{w}}(\widetilde{\mathbf{a}})}\alpha_{\widetilde{w}}
r_j(\widetilde{\mathbf{a}})-\sum_{j\in\mathcal{N}_{\widetilde{w}}(\widetilde{\mathbf{a}}_{-i})}
\alpha_{\widetilde{w}}
r_j(\widetilde{\mathbf{a}}_{-i})-\big(\sum_{j\in\mathcal{N}_{{w}^*}({\mathbf{a}}^*)}
\alpha_{{w}^*} r_j({\mathbf{a}^*})-
\sum_{j\in\mathcal{N}_{{w}^*}({\mathbf{a}^*_{-i}})
}\alpha_{{w}^*} r_j(\mathbf{a}^*_{-i})\big)\nonumber\\
&\stackrel{(b)}=\big(\hspace{-0.2cm}\sum_{j\in\mathcal{N}_{\widetilde{w}}(\widetilde{\mathbf{a}})}\hspace{-0.2cm}\alpha_{\widetilde{w}}
r_j(\widetilde{\mathbf{a}})+\hspace{-0.2cm}\sum_{j\in\mathcal{N}_{{w}^*}(\widetilde{\mathbf{a}}_{-i})
}\alpha_{{w}^*} r_j(\widetilde{\mathbf{a}}_{-i})\big)-
\big(\hspace{-0.2cm}\sum_{j\in\mathcal{N}_{{w}^*}({\mathbf{a}}^*)}\alpha_{{w}^*}
r_j({{\mathbf{a}}^*})+
\sum_{j\in\mathcal{N}_{\widetilde{w}}({\mathbf{a}}^*_{-i})}\alpha_{\widetilde{w}}
r_j({\mathbf{a}}^*_{-i})\big)\nonumber\\
&\stackrel{(c)}=\sum_{w\in\mathcal{W}}\alpha_w
\left(R_w(\widetilde{\mathbf{a}})-R_w({\mathbf{a}}^*)\right)
=R(\widetilde{\mathbf{a}})-R({\mathbf{a}}^*)\label{eqPotentialProperty}
\end{align}}
where $(a)$ is from the definition of the utility function
\eqref{eqUtilitySimple} ; $(b)$ is due to the fact that
$\mathbf{a}^*_{-i}=\widetilde{\mathbf{a}}_{-i}$; $(c)$ is due to
$\mathcal{N}_{{w}^*}(\widetilde{\mathbf{a}}_{-i})=\mathcal{N}_{{w}^*}(\widetilde{\mathbf{a}})$,
$\mathcal{N}_{\widetilde{w}}(\widetilde{\mathbf{a}}_{-i})=\mathcal{N}_{\widetilde{w}}({\mathbf{a}}^*)$,
and $R_{{w}}(\widetilde{\mathbf{a}})=R_{{w}}({\mathbf{a}}^*)$,
$\forall~{w}\ne\widetilde{w},~{w}^*$. From the assumption, user $i$
prefers to switch to BS $\widetilde{w}$, then
$U_i(\widetilde{\mathbf{a}})>U_i({\mathbf{a}}^*)$. This combined
with \eqref{eqPotentialProperty} yields
$R(\widetilde{\mathbf{a}})>R({\mathbf{a}}^*)$, which is a
contradiction to the optimality of $\ba^*$. Then we have
${\mathbf{a}}^*\in\arg\max_{\mathbf{a}}R(\mathbf{a})$ is a NE for
game $\mathcal{G}$. \hfill $\blacksquare$

\vspace{-0.2cm}
\subsection{Proof of Theorem \ref{theoremAppRatio}}
We show that \eqref{eqDefSubmodularity4} is true for all
$k\in\mathcal{Q}$ and $k\in\overline{\mathcal{Q}}$.

{\bf Step 1)} We first argue that for all $k\in\mathcal{Q}$,
\eqref{eqDefSubmodularity4} is true. When
$\mathbf{h}[k]\le\mathbf{m}[k]=\mathbf{g}[k]$,
\eqref{eqDefSubmodularity4} is trivially true as both sides of it
evaluate to zero. We then focus on the case
$\mathbf{h}[k]>\mathbf{m}[k]=\mathbf{g}[k]$.

Assuming $\mathbf{h}[k]>\mathbf{m}[k]=\mathbf{g}[k]$, we have
that{\small
\begin{align}
\mathbf{g}\vee\mathbf{e}_k\mathbf{h}[k]=\mathbf{m}\vee\mathbf{e}_k\mathbf{h}[k]=\mathbf{g}+c\times\mathbf{e}_k
\nonumber
\end{align}}
for some constant $c\ge 0$. Thus, for all $k\in\mathcal{Q}$ and
$\mathbf{h}[k]>\mathbf{m}[k]=\mathbf{g}[k]$, to show the inequality
\eqref{eqDefSubmodularity4}, it suffices to show the following {\it
decreasing difference} property{\small
\begin{align}
R_w(\mathbf{g}+\delta\times\mathbf{e}_k)-R_w(\mathbf{g})\le
R_w(\mathbf{m}+\delta\times\mathbf{e}_k)-R_w(\mathbf{m}),~\forall~\delta\ge
0, ~\textrm{and}~\mathbf{g}\ge\mathbf{m}\ge
0.\label{eqDefSubmodularityDelta}
\end{align}}
From \cite{topkis98}, we know that whenever the function
$R_w(\mathbf{x})$ is differentiable with respect to $\mathbf{x}[k]$,
the decreasing difference property of
\eqref{eqDefSubmodularityDelta} is equivalent to the following
property
{\small
\begin{align}
\lim_{\delta\to
0}\frac{R_w(\mathbf{g}+\delta\times\mathbf{e}_k)-R_w(\mathbf{g})}{\delta}-
\frac{R_w(\mathbf{m}+\delta\times\mathbf{e}_k)-R_w(\mathbf{m})}{\delta}\le
0, \ \forall~\mathbf{g}\ge\mathbf{m}\ge
0\label{eqDecreasingDerivativeLimit}.
\end{align}}
In what follows, we prove that for any $k$ with
$\mathbf{h}[k]>\mathbf{m}[k]=\mathbf{g}[k]$, the limit in
\eqref{eqDecreasingDerivativeLimit} exists and is non-positive. To
this end, a closer look at the function $R_w(\cdot)$ is necessary.
Let ${p}^k_{\mathbf{g}}$ denote the power allocation for channel $k$
when the best channel gain vector is $\mathbf{g}$, and let
$\lambda_{\mathbf{g}}$ denote the corresponding dual variable. From
the CAPA strategy, we have that
${p}^k_{\mathbf{g}}=[\lambda_{\mathbf{g}}-\frac{1}{\mathbf{g}[k]}]^+$.
Define the {\it active channel set} as
$\mathcal{K}_{\mathbf{g}}\triangleq\{k|
\lambda_{\mathbf{g}}-\frac{1}{\mathbf{g}[k]}\ge 0 \}$. From the fact
that the power constraint must be active for the CAPA strategy, we
have that {\small $
\sum_{k=1}^{K}p^k_{\mathbf{g}}=\sum_{k\in\mathcal{K}_{\mathbf{g}}}
\lambda_{\mathbf{g}}-\frac{1}{\mathbf{g}[k]}=\bar{p}_w $}, which
implies that{\small
\begin{align}
\lambda_{\mathbf{g}}=\frac{1}{|\mathcal{K}_{\mathbf{g}}|}
\left(\bar{p}_w+\sum_{k\in\mathcal{K}_{\mathbf{g}}}\frac{1}{\mathbf{g}[k]}\right)
\label{eqLambda}.
\end{align}}
Using this expression as well as the expression for
${p}^k_{\mathbf{g}}$, we have that{\small
\begin{align}
R_w(\mathbf{g})&=|\mathcal{K}_{\mathbf{g}}|
\log\left(\frac{\bar{p}_w+\sum_{k\in\mathcal{K}_{\mathbf{g}}}\frac{1}{\mathbf{g}[k]}}
{|\mathcal{K}_{\mathbf{g}}|}\right)+
\sum_{k\in\mathcal{K}_{\mathbf{g}}}\log(\mathbf{g}[k])=
|\mathcal{K}_{\mathbf{g}}|\log\left(\lambda_{\mathbf{g}}\right)+
\sum_{k\in\mathcal{K}_{\mathbf{g}}}\log(\mathbf{g}[k])\label{eqRateExpression}.
\end{align}}
We argue that when {\small$\mathbf{m}\le\mathbf{g}$}, we must have
{\small$\lambda_{\mathbf{g}}\le \lambda_{\mathbf{m}}$}. Otherwise,
if {\small$\lambda_{\mathbf{g}}>\lambda_{\mathbf{m}}$}, due to the
fact that
{\small$\frac{1}{\mathbf{g}[k]}\le\frac{1}{\mathbf{m}[k]}$}, we must
have
{\small${p}^k_{\mathbf{g}}>{p}^k_{\mathbf{m}},\forall~k\in\mathcal{K}_{\mathbf{m}}$},
which implies{\small
$\sum_{k\in\mathcal{K}_{\mathbf{m}}}\mathbf{p}^k_{\mathbf{g}}>
\sum_{k\in\mathcal{K}_{\mathbf{m}}}\mathbf{p}^k_{\mathbf{m}}=\bar{p}_w$},
a violation of the total power constraint.

Take any channel $k^*$ with
$\mathbf{h}[k^*]>\mathbf{m}[k^*]=\mathbf{g}[k^*]$, and define the
best channel gains {\it after} the increase on channel $k^*$ as
$\mathbf{m}^*=\mathbf{m}+\mathbf{e}_{k^*}\times\delta$ and
$\mathbf{g}^*=\mathbf{g}+\mathbf{e}_{k^*}\times \delta$,
respectively. Let $\mathcal{K}_{\mathbf{m}^*}$ and
$\mathcal{K}_{\mathbf{g}^*}$ denote the set of active channels.
Comparing $\mathcal{K}_{\mathbf{m}}$ and
$\mathcal{K}_{\mathbf{m}^*}$, we have the following four cases: m1)
there exists an $\epsilon>0$ such that for all $0\le\delta\le
\epsilon$, $\mathcal{K}_{\mathbf{m}}=\mathcal{K}_{\mathbf{m}^*}$;
m2) for all $\delta>0$,
$\mathcal{K}_{\mathbf{m}}\supset\mathcal{K}_{\mathbf{m}^*}$; m3) for
all $\delta>0$,
$\mathcal{K}_{\mathbf{m}}\subset\mathcal{K}_{\mathbf{m}^*}$; m4) for
all $\delta>0$,
$\mathcal{K}_{\mathbf{m}}\ne\mathcal{K}_{\mathbf{m}^*}$. Similarly,
we have four cases g1)--g4) comparing the sets
$\mathcal{K}_{\mathbf{g}}$ and $\mathcal{K}_{\mathbf{g}^*}$. In the
following we give the expression for  $\lim_{\delta\to
0}\frac{R_w(\mathbf{m}+\delta\times\mathbf{e}_k)-R_w(\mathbf{m})}{\delta}$
for each of the cases m1)--m4).


We first consider case m1). In the neighborhood of $0<\delta\le
\epsilon$,  $R_w(\mathbf{m}^*)$ can be expressed as{\small
\begin{align}
R_w(\mathbf{m}^*)=|\mathcal{K}_{\mathbf{m}}|\log\bigg(\frac{\bar{p}_w+
\sum_{k\in\mathcal{K}_{\mathbf{m}}}\frac{1}{\mathbf{m}^*[k]}}{|\mathcal{K}_{\mathbf{m}}|}\bigg)+
\sum_{k\in\mathcal{K}_{\mathbf{m}}}\log(\mathbf{m}^*[k]),~\forall~0\le\delta\le\epsilon.\nonumber
\end{align}}
Consequently, in the neighborhood of $0<\delta\le \epsilon$, we have
{\small
\begin{align}
\lim_{\delta\to
0^+}\frac{R_w(\mathbf{m}^*)-R_w(\mathbf{m})}{\delta}&=-\frac{|\mathcal{K}_{\mathbf{m}}|(\frac{1}{\mathbf{m}[k^*]})^2}
{\bar{p}_w+\sum_{k\in\mathcal{K}_{\mathbf{m}}}\frac{1}{\mathbf{m}[k]}}+\frac{1}{\mathbf{m}[k^*]}=
-\frac{1}
{\lambda_{\mathbf{m}}}{\frac{1}{(\mathbf{m}[k^*])^2}}+\frac{1}{\mathbf{m}[k^*]}.
\end{align}}
We then consider case m2). This case is shown by $4$ steps.

{\it Step (m2-1)}: We first show $k^*\in\mathcal{K}_{\mathbf{m}}$.
If on the contrary, $\lambda_{\mathbf{m}}-\frac{1}{\mathbf{m}[k^*]}<
0$, then due to the continuity of ${p}^{k^*}_{\mathbf{m}}$ with
respect to $\mathbf{m}$, there must exist an $\epsilon>0$ such that
for all $0<\delta<\epsilon$, $p^{k^*}_{\mathbf{m}^*}<0$, which is
equivalent to $\lambda_{\mathbf{m}^*}-\frac{1}{\mathbf{m}^*[k^*]}<
0$. This implies
$\mathcal{K}_{\mathbf{m}}=\mathcal{K}_{\mathbf{m}^*}$ for
$0<\delta<\epsilon$, which contradicts the assumption that
$\mathcal{K}_{\mathbf{m}}\supset\mathcal{K}_{\mathbf{m}^*},~\forall~\delta>0$.

{\it Step (m2-2)} We then argue that
{$k^*\in\mathcal{K}_{\mathbf{m}^*}$}. Assume the contrary, then {
$\lambda_{\mathbf{m}^*}-\frac{1}{\mathbf{m}^*[k^*]}<0$} . From the
previous step, we see that
$\lambda_{\mathbf{m}}-\frac{1}{\mathbf{m}[k^*]}\ge 0$. Then it must
be the case that $\lambda_{\mathbf{m}}>\lambda_{\mathbf{m}^*}$. Due
to the fact that for all other channels ${k}\ne{k}^*$,
{$\mathbf{m}[k]=\mathbf{m}^*[k]$}, then we must have
{$\bar{p}_w=\sum_{k=1}^{K}\mathbf{p}_{\mathbf{m}}^k>\sum_{k=1}^{K}\mathbf{p}_{\mathbf{m}^*}^k=\bar{p}_w$},
a contradiction.

{\it Step (m2-3)} We have argued that $k^*$ must remain in the
active set. Then for $\epsilon$ small enough, there must exist a
single channel $\tilde{k}\ne k^*$ such that
{\small$\tilde{k}\in\mathcal{K}_{\mathbf{m}}$} but {\small
$\tilde{k}\notin \mathcal{K}_{\mathbf{m}^*}$}, for all $0\le
\delta\le\epsilon$\footnote{If for all $\delta>0$, multiple channels
leave $\mathcal{K}_m$, then they must have the same magnitude--a
probability $0$ event. Our argument can also be applied to this
degenerate case, but for the sake of notational simplicity, we only
present the single $\tilde{k}$ case.}. The dual variables
$\lambda_{\mathbf{m}}$ and $\lambda_{\mathbf{m}^*}$ can be expressed
as{\small
\begin{align}
\lambda_{\mathbf{m}}&=\frac{1}{|\mathcal{K}_{\mathbf{m}}|}
\left(\bar{p}_w+\sum_{k\in\mathcal{K}_{\mathbf{m}}}\frac{1}{\mathbf{m}[k]}\right),~~
\lambda_{\mathbf{m}^*}=\frac{1}{|\mathcal{K}_{\mathbf{m}}|-1}
\bigg(\bar{p}_w+\sum_{k\in\mathcal{K}_{\mathbf{m}}\setminus\{\tilde{k},k^*\}}\frac{1}{\mathbf{m}[k]}+
\frac{1}{\mathbf{m}[k^*]+\delta}\bigg).
\end{align}}
The difference between the above two dual variables is given
by{\small
\begin{align}
0\stackrel{(a)}\le
\lambda_{\mathbf{m}}-\lambda_{\mathbf{m}^*}
&=\frac{-\lambda_{\mathbf{m}}+\frac{1}{\mathbf{m}\left[\tilde{k}\right]}
+\frac{1}{\mathbf{m}\left[{k}^*\right]}-\frac{1}{\mathbf{m}\left[{k^*}\right]+\delta}}
{|\mathcal{K}_{\mathbf{m}}|-1}\label{eqDifferenceLambda}
\end{align}}
where $(a)$ is from the fact that $\mathbf{m}^*\ge\mathbf{m}$, and
use the same argument in the paragraph following
\eqref{eqRateExpression}. Note that {\small
$\tilde{k}\in\mathcal{K}_{\mathbf{m}}$}, then
{\small$\lambda_{\mathbf{m}}-\frac{1}{\mathbf{m}[\tilde{k}]}\ge 0$}.
Combine this with \eqref{eqDifferenceLambda}, we have that {\small
$\frac{1}{\mathbf{m}[k^*]}-\frac{1}{\mathbf{m}[k^*]+\delta}\ge
\lambda_{\mathbf{m}}-\frac{1}{\mathbf{m}[\tilde{k}]}\ge 0$} for
arbitrary small $\delta>0$. Then it must be true that {\small
$\lambda_{\mathbf{m}}-\frac{1}{\mathbf{m}[\tilde{k}]}=0$}.

{\it Step (m2-4)} Using the result obtained in Step (m2-3) and the
rate expression \eqref{eqRateExpression}, we can express the
difference of the rate $R_w(\mathbf{m}^*)$ and $R_w(\mathbf{m})$
as{\small
\begin{align}
R_w(\mathbf{m}^*)-R_w(\mathbf{m})
&\stackrel{(a)}=(|\mathcal{K}_{\mathbf{m}}|-1)\log\bigg(\frac{(\bar{p}_w+\sum_{k\in\mathcal{K}_{\mathbf{m}}\setminus\{\tilde{k},k^*\}}\frac{1}{\mathbf{m}[k]}+
\frac{1}{\mathbf{m}[k^*]+\delta})|\mathcal{K}_\mathbf{m}|\mathbf{m}\big[\tilde{k}\big]}
{(|\mathcal{K}_{\mathbf{m}}|-1)|\mathcal{K}_\mathbf{m}|}\bigg)+
\log\bigg(\frac{\mathbf{m}[k^*]+\delta}{\mathbf{m}[k^*]}\bigg)\nonumber\\
&\stackrel{(b)}=(|\mathcal{K}_{\mathbf{m}}|-1)\log\bigg(\frac{|\mathcal{K}_{\mathbf{m}}|+
\frac{\mathbf{m}(\tilde{k})}{\mathbf{m}[{k}^*]+\delta}-\frac{\mathbf{m}\left[\tilde{k}\right]}{\mathbf{m}[{k}^*]}-1}
{|\mathcal{K}_{\mathbf{m}}|-1}\bigg)+
\log\bigg(\frac{\mathbf{m}[k^*]+\delta}{\mathbf{m}[k^*]}\bigg)\nonumber\\
&=(|\mathcal{K}_{\mathbf{m}}|-1)\log\bigg(1+\frac{
\frac{\mathbf{m}\left[\tilde{k}\right]}{\mathbf{m}\left[{k}^*\right]+\delta}
-\frac{\mathbf{m}\left[\tilde{k}\right]}{\mathbf{m}[{k}^*]}}
{|\mathcal{K}_{\mathbf{m}}|-1}\bigg)+
\log\bigg(\frac{\mathbf{m}[k^*]+\delta}{\mathbf{m}[k^*]}\bigg)
\end{align}}
where in $(a),(b)$ we have used the fact that
$\lambda_{\mathbf{m}}=\frac{1}{\mathbf{m}[\tilde{h}]}$. Using
L'Hopital's rule, we obtain{\small
\begin{align}
\lim_{\delta\to
0^+}\frac{R_w(\mathbf{m}^*)-R_w(\mathbf{m})}{\delta}&=\lim_{\delta\to
0^+}\frac{-
\frac{\mathbf{m}[\tilde{k}]}{(\mathbf{m}[k^*]+\delta)^2}}{1+\frac{1}{|\mathcal{K}_{\mathbf{m}}|-1}
\frac{\mathbf{m}[\tilde{k}]}{\mathbf{m}[{k}^*]+\delta}-\frac{\mathbf{m}[\tilde{k}]}{\mathbf{m}[{k}^*]}}+
\frac{1}{(\mathbf{m}[k^*]+\delta)}\nonumber\\
&=-\frac{\mathbf{m}[\tilde{k}]}{(\mathbf{m}[k^*])^2}+\frac{1}{\mathbf{m}[k^*]}=
-\frac{1}{\lambda_{\mathbf{m}}}\frac{1}{(\mathbf{m}[k^*])^2}+\frac{1}{\mathbf{m}[k^*]}.\label{eqCaseH2}
\end{align}}
For the cases m3)--m4), the derivation is similar to the cases of
m1)-- m2). The key observation is still  that the channel ${k}^*$
must satisfy $k^*\in\mathcal{K}_{\mathbf{m}}$ and
$k^*\in\mathcal{K}_{\mathbf{m}^*}$, and that the channel $\tilde{k}$
that leaves or joins the set $\mathcal{K}_{\mathbf{m}^*}$ must
satisfy $\lambda_{\mathbf{m}}=\frac{1}{\mathbf{m}[\tilde{k}]}$.
For these cases, \eqref{eqCaseH2} again holds true. 

Fix $\delta<0$, and redo the above analysis by switching the role of
$\mathbf{m}$ and $\mathbf{m}^*$ for all four possible cases, we can
obtain {\small $\lim_{\delta\to
0^-}\frac{R_w(\mathbf{m}^*)-R_w(\mathbf{m})}{\delta}=
-\frac{1}{\lambda_{\mathbf{m}}}\frac{1}{(\mathbf{m}[k^*])^2}+\frac{1}{\mathbf{m}[k^*]}.
$} Consequently, we have that for all $k^*$ that satisfies
$\mathbf{h}[k^*]>\mathbf{m}[k^*]=\mathbf{g}[k^*]$, the following is
true{\small
\begin{align}
\lim_{\delta\to
0}\frac{R_w(\mathbf{m}+\mathbf{e}_{k^*}\times\delta)-R_w(\mathbf{m})}{\delta}&=
-\frac{1}{\lambda_{\mathbf{m}}}\frac{1}{(\mathbf{m}[k^*])^2}+\frac{1}{\mathbf{m}[k^*]},~\forall~\mathbf{g}\ge\mathbf{m}\ge
0.
\end{align}}
For case g1)-g4), the exact same argument leads to the same result.
In summary, we obtain{\small
\begin{align}
&\lim_{\delta\to
0}\frac{R_w(\mathbf{g}^*)-R_w(\mathbf{g})}{\delta}-\frac{R_w(\mathbf{m}^*)-R_w(\mathbf{m})}{\delta}
=-\frac{1}{\lambda_{\mathbf{g}}}\frac{1}{(\mathbf{g}[k^*])^2}+\frac{1}{\mathbf{g}[k^*]}+
\frac{1}{\lambda_{\mathbf{m}}}\frac{1}{(\mathbf{m}[k^*])^2}-\frac{1}{\mathbf{m}[k^*]}.
\end{align}}
Recall that ${k}^* \in\mathcal{Q}$, which means that
$\mathbf{m}[k^*]=\mathbf{g}[k^*]$. Using the fact that
$\mathbf{g}\ge\mathbf{m}$, and
$\lambda_{\mathbf{g}}\le\lambda_\mathbf{m}$, we conclude that
\eqref{eqDecreasingDerivativeLimit} is true for all $k$ with
$\mathbf{h}[k]>\mathbf{m}[k]=\mathbf{g}[k]$.

{\bf Step 2)} We then argue that for any channel
$k\in\overline{\mathcal{Q}}$, \eqref{eqDefSubmodularity4} must be
true. For any $\mathbf{g}\ge \mathbf{m}\ge 0$, pick
$k\in\overline{\mathcal{Q}}$, we have the following three cases: 1)
$\mathbf{h}[k]\le \mathbf{m}[k]$; 2) $\mathbf{m}[k]< \mathbf{h}[k]<
\mathbf{g}[k]$; 3) $\mathbf{h}[k]\ge\mathbf{g}[k]$. 
Verifying case 1)--case 2) is straightforward. 
For case 3) we have{\small
\begin{align}
R_w(\mathbf{m}\vee \mathbf{e}_k\mathbf{h}[k])-R_w(\mathbf{m})
&=R_w(\mathbf{m}\vee \mathbf{e}_k\mathbf{h}[k])-R_w(\mathbf{m}\vee
\mathbf{e}_k\mathbf{g}[k])+R_w(\mathbf{m}\vee
\mathbf{e}_k\mathbf{g}[k])-R_w(\mathbf{m})\nonumber\\
&\ge R_w(\mathbf{m}\vee
\mathbf{e}_k\mathbf{h}[k])-R_w(\mathbf{m}\vee
\mathbf{e}_k\mathbf{g}[k])
\end{align}}
where the inequality is due to the monotonicity property. It is
sufficient to show{\small
\begin{align}
R_w(\mathbf{g}\vee \mathbf{e}_k\mathbf{h}[k])-R_w(\mathbf{g})\le
R_w(\mathbf{m}\vee \mathbf{e}_k\mathbf{h}[k])-R_w(\mathbf{m}\vee
\mathbf{e}_k\mathbf{g}[k]), ~\forall~\mathbf{h}\ge
0,~\mathbf{g}\ge\mathbf{m}\ge 0\label{eqCase3}.
\end{align}}
Let $\widetilde{\mathbf{m}}=\mathbf{m}\vee\mathbf{e}_k\mathbf{g}[k]$
and $\mathbf{h}=\mathbf{g}+\delta_k\mathbf{e}_k$, for some
$\delta_k>0$. Clearly $\widetilde{\mathbf{m}}[k]=\mathbf{g}[k]$.
Then to show \eqref{eqCase3}, it is sufficient to show that for all
$k$ such that $\widetilde{\mathbf{m}}[k]=\mathbf{g}[k]$, we have
{\small
\begin{align}
R_w(\mathbf{g}+\mathbf{e}_k\delta_k)-R_w(\mathbf{g})\le
R_w(\widetilde{\mathbf{m}}+
\mathbf{e}_k\delta_k)-R_w(\widetilde{\mathbf{m}}), \
\forall~\delta\ge0, \mathbf{g}\ge\widetilde{\mathbf{m}}\ge 0
\end{align}}
which reduces to the case in Step 1) (cf. condition
\eqref{eqDefSubmodularityDelta}). We also have that
$\eqref{eqDefSubmodularity4}$ is true.

Combining with our argument in Step 1), we conclude that
\eqref{eqDefSubmodularity4} is true for all $k\in\mathcal{K}$.

\vspace{-0.3cm} \subsection{Proof of Theorem
\ref{theoremConvergence}}\label{appProof}

Let $\mathbf{c}^{(t)}$ denote the {\it better-reply association} at
time $t$: $\mathbf{c}^{(t)}[i]=w_i^{*(t)}$. Define two sets
$\mathcal{C}$ and $\mathcal{A}$: $ \mathbf{c}\in \mathcal{C}
\Rightarrow \mathbf{c} \textrm{~~appears infinitely often (i.o.) in
~} \{\mathbf{c}^{(t)}\}_{t=1}^{\infty}$, and $ \mathbf{a}\in
\mathcal{A} \Rightarrow \mathbf{a} \textrm{~~i.o. in ~}
\{\mathbf{a}^{(t)}\}_{t=1}^{\infty}$.

The first claim is that there exist $\mathbf{a}^*\in \mathcal{A}$
that is a pure NE for game $\mathcal{G}$. Observe that the sets
$|\mathcal{A}|>0$ and $|\mathcal{C}|>0$ due to the finiteness of the
possible association profiles. Suppose $|\mathcal{A}|=1$, then the
single element in $\mathcal{A}$, say $\mathbf{a}^*$, must be a NE.
Suppose $|\mathcal{A}|>1$, and choose $\mathbf{c}^*\in\mathcal{C}$.
Pick a time $t$ such that $\mathbf{c}^{(t)}=\mathbf{c}^*$. Note that
$\mathbf{c}^*[i]$ is in the front of the memory for each user $i$,
then with probability at least $(\frac{1}{M})^N$,
$\mathbf{a}^{(t+1)}=\mathbf{c}^*$. This implies
$\mathbf{c}^*\in\mathcal{A}$. If $\mathbf{c}^*$ is a NE, then our
claim is proved. If $\mathbf{c}^*$ is not a NE, we will show that
with positive probability, we can construct a finite sequence that
leads to a NE. To this end, consider the following steps of
operation.

{\bf Step 1)}: With probability at least $(\frac{1}{M})^N$,
$\mathbf{a}^{(t+1)}=\mathbf{c}^*$. Because $\mathbf{c}^*$ is not a
NE, then there exists an $i\in\mathcal{N}$ such that
$\mathbf{c}^{(t+1)}[i]\ne\mathbf{c}^{(t)}[i]$. Similarly as in the
proof of Theorem 3, we can show that
${R}(\mathbf{a}^{t+1})<{R}\left([\mathbf{c}^{(t+1)}[i],\mathbf{a}_{-i}^{(t+1)}]\right)$.
With probability at least $(\frac{1}{M})^N$, every user $j\ne i$
samples $\mathbf{c}^{(t)}[j]$, which is now at the second slot in
the memory, while user $i$ samples $\mathbf{c}^{(t+1)}[i]$. This
event leads to
$\mathbf{a}^{(t+2)}=[\mathbf{c}^{(t+1)}[i],\mathbf{a}_{-i}^{(t+1)}]$,
and we have ${R}(\mathbf{a}^{(t+2)})>{R}(\mathbf{a}^{(t+1)})$. Put
index $i$ into a set $\mathcal{U}$ : $\mathcal{U}=\{i\}$. Note in
this
stage, we have: $\mathbf{a}^{(t+2)}[i]=\mathbf{c}^{(t+1)}[i]$. 
Continue this process, until we reach a time $t+n\le t+N$ such that
only users in the set $\mathcal{U}$ are willing to switch, i.e.,
$\forall \ j\in\mathcal{E}$, $\mathbf{c}^{(T)}_j=\mathbf{a}^{(T)}_j$. 
Note that the requirement $M\ge N$ ensures that for all $i$, the set
of best responses $\{\mathbf{c}^{(m)}[i]\}_{m=t}^{(t+n)}$ is still
in user $i$'s memory. Let $T=t+n$. Let
$\mathcal{E}=\mathcal{N}\setminus\mathcal{U}$.

{\bf Step 2)}: Observe that for all $i\in\mathcal{U}$, there must
exist a constant $k_i$ such that $0<k_i<n\le N$ and that its current
association $\mathbf{a}^{(T)}[i]$ is sampled from its $k_i$th
memory, i.e., $\mathbf{c}^{(T-k_i)}[i]=\mathbf{a}^{(T)}[i]$. Pick
$q\in\mathcal{U}$ that has the largest $k_i$ and is willing to
switch at time $T$: {\small$q=\arg\max_{i\in\mathcal{U},
\mathbf{c}^{(T)}[i]\ne\mathbf{a}^{(T)}[i]}k_i$}. We can now shift
$\mathbf{c}^{(T-N)}$ out of the memory and still be able to
construct
$\mathbf{a}^{(T+1)}=[\mathbf{c}^{(T)}[q],\mathbf{a}_{-q}^{(T)}]$
with positive probability, because all the elements in
$\mathbf{a}_{-q}^{(T)}$ must have been appeared once in
$\{\mathbf{c}^{(t)}\}^{T}_{t=T-N+1}$. Move $q$ out of $\mathcal{U}$
and into $\mathcal{E}$, let $T=T+1$ and continue Step 2) until only
users in the set $\mathcal{E}$ are willing to switch. Change the
role of $\mathcal{U}$ and $\mathcal{E}$, and continue Step 2).

Repeating Step 2), we construct a sequence
$\{R(\mathbf{a}^{(t+l)})\}$ that is strictly increasing. Due to the
finiteness of the choice of $\mathbf{a}$, there must exist a {\it
finite} time instance $T^*$ after which it is not possible to find
an association that differs from $\mathbf{a}^{(t+T^*)}$ with a
single element and still have strict better system throughput.
Consequently, $\mathbf{a}^*=\mathbf{a}^{(t+T^*)}$ is an equilibrium
profile. Thus, with {\it positive} probability, a NE profile
$\mathbf{a}^*$ appears after $\mathbf{a}^{(t+1)}$ in {\it finite}
steps. Because $\mathbf{a}^{(t+1)}=\mathbf{c}^*$, with
$\mathbf{c}^*$ happens i.o., we must also have $\mathbf{a}^*$ i.o.,
that is, $\mathbf{a}^*\in\mathcal{A}$. The claim is proved.

The next claim is that the algorithm converges to $\mathbf{a}^*$
with probability 1. Let $\{t_k\}_{k=1}^{\infty}$ denote the
subsequence of $\{t\}$ in which $\mathbf{a}^*$ happens. Define the
event: $
C_k\triangleq\bigcap_{l=1}^{M}\{\mathbf{a}^{(t_k+l)}=\mathbf{a}^*\}$,
that is, starting from a time $t_k$, $\mathbf{a}^*$ appears $M+1$
times consecutively. When $C_k$ happens, we have: 1) at time
$t_k+M+1$, $\mathbf{c}^{(t_k+M+1)}=\mathbf{a}^*$ because
$BR_i(\mathbf{a}^*)=\mathbf{a}^*[i]$ $\forall~i$; 2)
$\mathbf{a}^{(t_k+M+l)}=\mathbf{a}^*$ for all $l\ge 1$ because after
time $(t_k+M+l)$, each user $i$'s memory will solely consist of
$\mathbf{a}^*[i]$. Note that if $\mathbf{a}^{(t_k)}=\mathbf{a}^*$
appears, with probability at least $(\frac{1}{M})^N$,
$\mathbf{a}^{(t_k+1)}=\mathbf{a}^*$. This implies $ {\rm
Pr}(C_k)\ge(\frac{1}{M})^{N\times M}$. Let $C^c_k$ denote the
complement set of $C_k$. We have:{\small
\begin{align}
\hspace{-0.1cm}&{\rm Pr}\hspace{-0.1cm}\left(\bigcap_{t\ge
1}C^c_k\right)\hspace{-0.15cm}=\hspace{-0.15cm}\lim_{T\to\infty}\hspace{-0.15cm}{\rm
Pr}\hspace{-0.1cm}\left(\bigcap_{t=1}^{
T}C^c_k\right)\hspace{-0.1cm}=\hspace{-0.1cm}\lim_{T\to\infty}\prod_{k=1}^{T-1}\hspace{-0.15cm}
\left(1-{\rm Pr}(C_k)\right)
\hspace{-0.1cm}\le\lim_{T\to\infty}\Big(1-(\frac{1}{M})^{N\times
M}\Big)^{T-1}\hspace{-0.3cm}=0.
\end{align}}
This says $ {\rm Pr}(\mathbf{a}^{(t)} ~\textrm{converges to some }
\mathbf{a}^*\in\mathcal{A}^*~\textrm{eventually})=1.
$\hfill${\blacksquare}$

{ \small \vspace{-0.2cm}
\bibliographystyle{IEEEbib}
\bibliography{ref}

\begin{thebibliography}{10}

\bibitem{McNair04}
J.~McNair and F.~Zhu,
\newblock ``Vertical handoffs in fourth-generation multinetwork environments,''
\newblock {\em IEEE Wireless Communications}, vol. 11, no. 3, pp. 8 -- 15,
  2004.

\bibitem{yeh11}
S.-P. Yeh, S.~Talwar, G.~Wu, N.~Himayat, and K.~Johnsson,
\newblock ``Capacity and coverage enhancement in heterogeneous networks,''
\newblock {\em IEEE Wireless Communications}, vol. 18, no. 3, pp. 32 --38, june
  2011.

\bibitem{fcc10}
``{IEEE} 802.22/d1.0, december 2010 draft standard for wireless regional area
  networks part 22: Cognitive wireless ran medium access control and physical
  layer specifications,'' Dec 2010.

\bibitem{kauffmann07}
B.~Kauffmann, F.~Baccelli, and A.~Chaintreau,
\newblock ``Measurement-based self organization of interfering 802.11 wireless
  access network,''
\newblock in {\em the Proceedings of IEEE INFOCOM}, 2007, pp. 1451--1459.

\bibitem{Kavitha12}
V.~Kavitha, E.~Altman, R.~El-Azouzi, and R.~Sundaresan,
\newblock ``Opportunistic scheduling in cellular systems in the presence of
  noncooperative mobiles,''
\newblock {\em IEEE Transactions on Information Theory}, vol. 58, no. 3, pp.
  1757 --1773, march 2012.

\bibitem{Bianchi07}
G.~Bianchi, A.~Di~Stefano, C.~Giaconia, L.~Scalia, G.~Terrazzino, and
  I.~Tinnirello,
\newblock ``Experimental assessment of the backoff behavior of commercial
  {IEEE} 802.11b network cards,''
\newblock in {\em IEEE INFOCOM 2007}, may 2007, pp. 1181 --1189.

\bibitem{Nuggehalli08}
P.~Nuggehalli, M.~Sarkar, K.~Kulkarni, and R.R. Rao,
\newblock ``A game-theoretic analysis of qos in wireless mac,''
\newblock in {\em IEEE INFOCOM 2008}, pp. 1903 --1911.

\bibitem{Kong07}
Zhen Kong, Yu-Kwong Kwok, and Jiangzhou Wang,
\newblock ``On game theoretic rate-maximizing packet scheduling in
  non-cooperative wireless networks,''
\newblock in {\em IEEE WoWMoM 2007}, june 2007, pp. 1 --4.

\bibitem{song05a}
G.~Song and Y.~Li,
\newblock ``Cross-layer optimization for {OFDM} wireless networks--part {I}:
  Theoretical framework,''
\newblock {\em IEEE Transactions on Wireless Communications}, vol. 4, no. 2,
  pp. 614--624, 2005.

\bibitem{song05b}
G.~Song and Y.~Li,
\newblock ``Cross-layer optimization for {OFDM} wireless networks--part {II}:
  Algorithm development,''
\newblock {\em IEEE Transactions on Wireless Communications}, vol. 4, no. 2,
  pp. 625--634, 2005.

\bibitem{tse_m}
D.~N.~C. Tse,
\newblock ``Optimal power allocation over parallel {G}aussinan broadcast
  channels,''
\newblock http://degas.eecs.berkeley.edu/~dtse.

\bibitem{han04}
Z.~Han, F.~R. Farrokhi, Z.~Ji, and K.J.R. Liu,
\newblock ``Optimal subchannel allocation scheme in multicell {OFDMA}
  systems,''
\newblock in {\em Proc. IEEE VTC}, 2004, pp. 1821--1825.

\bibitem{Li06}
G.~Li and H.~Liu,
\newblock ``Downlink radio resrouce allocation for multi-cell {OFDMA} system,''
\newblock {\em IEEE Transactions on Wireless Communications}, vol. 5, pp.
  3451--3459, 2006.

\bibitem{saraydar01}
C.~U. Sarayda, N.~B. Mandayam, and D.~J. Goodman,
\newblock ``Pricing and power control in a multicell wireless data network,''
\newblock {\em IEEE Journal on selected areas in communications}, vol. 19, no.
  10, pp. 1883--1892, 2001.

\bibitem{hong11_infocom}
M.~Hong, A.~Garcia, and J.~Barrera,
\newblock ``Joint distributed {AP} selection and power allocation in cognitive
  radio networks,''
\newblock in {\em the Proceedings of the IEEE INFOCOM}, 2011.

\bibitem{gao11}
L.~Gao, X.~Wang, G.~Sun, and Y.~Xu,
\newblock ``A game approach for cell selection and resource allocation in
  heterogeneous wireless networks,''
\newblock in {\em the Proceeding of the SECON}, 2011.

\bibitem{hong12_icassp}
M.~Hong and Z.-Q. Luo,
\newblock ``Joint linear precoder optimization and base station selection for
  an uplink {MIMO} network: A game theoretic approach,''
\newblock in {\em the Proceedings of the IEEE ICASSP}, 2012.

\bibitem{Yang07}
S.~Yang and B.~Hajek,
\newblock ``{VCG-K}elly mechanisms for allocation of divisible goods: adapting
  {VCG} mechanisms to one-dimensional signals,''
\newblock {\em IEEE Journal of Selected Areas of Communications}, vol. 25, pp.
  1237--1243, 2007.

\bibitem{jain10}
R.~Jain and J.~Walrand,
\newblock ``An efficient {N}ash-implementation mechanism for divisible resource
  allocation,''
\newblock {\em Automatica}, vol. 46, pp. 1276--1283, 2010.

\bibitem{Roughgarden00howbad}
T.~Roughgarden and E.~Tardos,
\newblock ``How bad is selfish routing?,''
\newblock {\em Journal of the ACM}, vol. 49, pp. 236--259, 2000.

\bibitem{Johari:2004}
R~Johari and J.~N. Tsitsiklis,
\newblock ``Efficiency loss in a network resource allocation game,''
\newblock {\em Math. Oper. Res.}, vol. 29, pp. 407--435, 2004.

\bibitem{vetta02}
A.~Vetta,
\newblock ``Nash equilibria in competitive societies, with applications to
  facility location, traffic routing and auctions,''
\newblock in {\em IEEE FOCS}, 2002.

\bibitem{ai08}
X.~Ai, V.~Srinivasan, and C.-K. Tham,
\newblock ``Optimality and complexity of pure nash equilibria in the coverage
  game,''
\newblock {\em IEEE Journal on Selected Areas in Communications}, vol. 26, no.
  7, pp. 1170 --1182, 2008.

\bibitem{Goemans06}
M.X. Goemans, L.~Li, V.S. Mirrokni, and M.~Thottan,
\newblock ``Market sharing games applied to content distribution in ad hoc
  networks,''
\newblock {\em IEEE J. Sel. Areas Commun.}, vol. 24, no. 5, pp. 1020 -- 1033,
  2006.

\bibitem{luo08a}
Z.-Q. Luo and S.~Zhang,
\newblock ``Dynamic spectrum management: Complexity and duality,''
\newblock {\em IEEE Journal of Selected Topics in Signal Processing}, vol. 2,
  no. 1, pp. 57--73, 2008.

\bibitem{Razaviyayn11Asilomar}
M.~Razaviyayn, M.~Hong, and Z.-Q. Luo,
\newblock ``Linear transceiver design for a {MIMO} interfering broadcast
  channel achieving max-min fairness,''
\newblock in {\em 2011 Asilomar Conference on Signals, Systems, and Computers},
  2011.

\bibitem{clark71}
E.~Clark,
\newblock ``Multipart pricing of public goods,''
\newblock {\em Public Choice}, vol. 2, pp. 19--33, 1971.

\bibitem{groves73}
T.~Groves,
\newblock ``Incentives in teams,''
\newblock {\em Econometrica}, vol. 41, pp. 617--631, 1973.

\bibitem{Vickrey61}
W.~Vickrey,
\newblock ``Counterspeculation, auctions, and competitive sealed tenders,''
\newblock {\em Journal of Finance}, pp. 8--37, 1961.

\bibitem{goldsmith05}
A.~Goldsmith,
\newblock {\em Wireless Communications},
\newblock Combridge University Press, New York, 2005.

\bibitem{borgers10}
T.~Borgers,
\newblock {\em An Introduction to the Theory of Mechanism Design},
\newblock www-personal.umich.edu/~tborgers/, 2010.

\bibitem{tse98part1}
D.~N.~C. Tse and S.~V. Hanly,
\newblock ``Multiaccess fading channels. {I}. polymatroid structure, optimal
  resource allocation and throughput capacities,''
\newblock {\em IEEE Trans. Inf. Theory}, vol. 44, no. 7, pp. 2796 --2815, 1998.

\bibitem{yu11}
W.~Yu, T.~Kwon, and C.~Shin,
\newblock ``Multicell coordination via joint scheduling beamforming and power
  spectrum adaptation,''
\newblock in {\em Proceedings of IEEE INFOCOM}, 2011.

\bibitem{rappaport02}
T.~S. Rappaport,
\newblock {\em Wireless communications principles and practices},
\newblock Prentice-Hall, 2002.

\bibitem{bu06}
T.~Bu, L.~Li, and R.~Ramjee,
\newblock ``Generalized proportional fair scheduling in third generation
  wireless data networks,''
\newblock in {\em Proceedings of IEEE INFOCOM 2006}, april 2006, pp. 1 --12.

\bibitem{benedict67}
T.~R. Benedict and T.~T. Soong,
\newblock ``The joint estimation of signal and noise from the sum evelope,''
\newblock {\em IEEE Transactions on Information Theory}, vol. 13, no. 3, pp.
  447--454, 1967.

\bibitem{pauluzzi00}
D.~R. Pauluzzi and N.~C. Beaulieu,
\newblock ``A comparison of {SNR} estimation techniques for the {AWGN}
  channel,''
\newblock {\em IEEE Transactions on Communications}, vol. 48, no. 10, pp.
  1681--1691, 2000.

\bibitem{hong11a}
M.~Hong and A.~Garcia,
\newblock ``Averaged iterative water-filling algorithm: Robustness and
  convergence,''
\newblock {\em IEEE Transactions on Signal Processing}, vol. 59, no. 5, pp.
  2448 --2454, may 2011.

\bibitem{topkis98}
D.M. Topkis,
\newblock {\em Supermodularity and complementarity},
\newblock Princeton University Press, 1998.

\end{thebibliography}
}


%

\end{document}